\documentclass[fleqn,usenatbib]{mnras}

\usepackage{newtxtext,newtxmath}

\usepackage[T1]{fontenc}
\usepackage{ae,aecompl}

\usepackage{float}

\usepackage{multirow}
\usepackage{subcaption}
\usepackage{booktabs}
\usepackage[dvipsnames]{xcolor}
\usepackage{soul}
\usepackage{graphicx}	
\usepackage{amsmath}	

\usepackage{amssymb}	
\usepackage{lineno}
\usepackage{xspace}
\usepackage{comment}

\usepackage{eso-pic}

\AddToShipoutPictureBG*{%
  \AtPageUpperLeft{%
    \hspace{0.75\paperwidth}%
    \raisebox{-3.5\baselineskip}{%
      \makebox[0pt][l]{\textnormal{DES-2019-0452}}
}}}%

\AddToShipoutPictureBG*{%
  \AtPageUpperLeft{%
    \hspace{0.75\paperwidth}%
    \raisebox{-4.5\baselineskip}{%
      \makebox[0pt][l]{\textnormal{FERMILAB-PUB-21-255-AE}}
}}}%

\newcommand{\code}[1]{\texttt{#1}\xspace}

\newcommand{\healpix}{\textsc{HEALPix}}
\newcommand{\healpy}{\textsc{healpy}\xspace}

\newcommand{\gold}{\code{Y3 GOLD}}
\newcommand{\threextwo}{3$\times$2pt }
\newcommand{\twoxtwo}{2$\times$2pt }
\newcommand{\redmagic}{\textsc{redMaGiC}\xspace}
\newcommand{\maglim}{\textsc{MagLim}\xspace}
\newcommand{\baosample}{\code{BAO SAMPLE}}
\newcommand{\cosmolike}{\textsc{CosmoLike}}
\newcommand{\cosmosis}{\textsc{CosmoSIS}}
\newcommand{\Camb}{\texttt {\textsc{Camb}}}
\newcommand{\Halofit}{{\texttt{Halofit}}}

\newcommand{\treecorr}{\textsc{TreeCorr}}
\newcommand{\balrog}{\textsc{Balrog}}
\newcommand{\dnf}{\code{DNF}}

\newcommand{\isd}{\textsc{ISD}\xspace}
\newcommand{\enet}{\textsc{ENet}\xspace}

\newcommand{\nside}{\ifmmode N_{\mathrm{side}}\else $N_{\mathrm{side}}$\fi}

\newcommand{\airmass}{\textit{airmass}}
\newcommand{\fwhm}{\textit{fwhm}}
\newcommand{\fwhmfluxrad}{\textit{fwhm\_fluxrad}}
\newcommand{\exptime}{\textit{exptime}}
\newcommand{\teff}{\textit{t\_eff}}
\newcommand{\teffexptime}{\textit{t\_eff\_exptime}}
\newcommand{\skybrite}{\textit{skybrite}}
\newcommand{\skyvar}{\textit{skyvar}}
\newcommand{\skyvarsqrt}{\textit{skyvar\_sqrt}}
\newcommand{\skyvaruncert}{\textit{skyvar\_uncertainty}}
\newcommand{\sigmamagzero}{\textit{sigma\_mag\_zero}}
\newcommand{\fgcmgry}{\textit{fgcm\_gry}}
\newcommand{\maglimdepth}{\textit{maglim}}
\newcommand{\sofdepth}{\textit{sof\_depth}}
\newcommand{\magautodepth}{\textit{magauto\_depth}}
\newcommand{\stars}{\textit{stars\_1620}}
\newcommand{\stellardens}{\textit{stellar\_dens}}
\newcommand{\sfd}{\textit{sfd98}}

\newcommand{\pcmap}[1]{\textit{pca}#1}

\title[]{Dark Energy Survey Year 3 Results: Galaxy clustering and systematics treatment for lens galaxy samples}

\author[DES Collaboration]{
\parbox{\textwidth}{
\Large
M.~Rodr\'{i}guez-Monroy,$^{1}$
N.~Weaverdyck,$^{2}$
J.~Elvin-Poole,$^{3,4}$
M.~Crocce,$^{5,6}$
A.~Carnero~Rosell,$^{7,8,9}$
F.~Andrade-Oliveira,$^{10,8}$
S.~Avila,$^{11}$
K.~Bechtol,$^{12}$
G.~M.~Bernstein,$^{13}$
J.~Blazek,$^{14,15}$
% ...
\begin{center} 
(DES Collaboration) 
\end{center}
}
}

\date{Accepted XXX. Received YYY; in original form ZZZ}

\pubyear{2021}

\begin{document}
\label{firstpage}
\pagerange{\pageref{firstpage}--\pageref{lastpage}}
\maketitle

\begin{abstract}
In this work we present the galaxy clustering measurements of the two DES lens galaxy samples: a magnitude-limited sample optimized for the measurement of cosmological parameters, \maglim, and a sample of luminous red galaxies selected with the \redmagic algorithm. \maglim \, / \redmagic sample contains over 10 million / 2.5 million galaxies and is divided into six / five photometric redshift bins spanning the range $z\in[0.20,1.05]$ / $z\in[0.15,0.90]$. Both samples cover 4143 $\deg^2$ over which we perform our analysis blind, measuring the angular correlation function with a S/N $\sim 63$ for both samples. In a companion paper \citep{y3-3x2ptkp}, these measurements of galaxy clustering are combined with the correlation functions of cosmic shear and galaxy-galaxy lensing of each sample to place cosmological constraints with a 3$\times$2pt analysis. We conduct a thorough study of the mitigation of systematic effects caused by the spatially varying survey properties and we correct the measurements to remove artificial clustering signals. We employ several decontamination methods with different configurations to ensure the robustness of our corrections and to determine the systematic uncertainty that needs to be considered for the final cosmology analyses. We validate our fiducial methodology using log-normal mocks, showing that our decontamination procedure induces biases no greater than $0.5\sigma$ in the $(\Omega_m, b)$ plane, where $b$ is galaxy bias. We demonstrate that failure to remove the artificial clustering would introduce strong biases up to $\sim 7 \sigma$ in $\Omega_m$ and of more than $4 \sigma$ in galaxy bias. 

\end{abstract}




\section{Introduction}\label{sec:introduction}
The current Standard Model of Cosmology, $\Lambda$CDM, provides an excellent fit to current observations, including distance measurements to Type Ia supernovae (SNIa) \citep{1998AJ....116.1009R, 1999ApJ...517..565P}, the cosmic microwave background (CMB) fluctuations \citep{2003ApJS..148..175S, 2020A&A...641A...6P} and the large-scale structure of the Universe \citep{2017MNRAS.470.2617A, 2019PhRvL.122q1301A, 2021PhRvD.103h3533A}, with only six free parameters. In addition, photometric galaxy surveys, such as the Kilo-Degree Survey \citep[KiDS, ][]{kids:2013}, Hyper Suprime-Cam Subaru Strategic Program \citep[HSC-SSP, ][]{hscssp:2018} and the Dark Energy Survey \citep[DES, ][]{des:2005} are now reaching a level of sensitivity that competes with the most precise determinations of cosmological parameters currently available \citep{DESY1cosmo:2018, Heymans:2021}. The comparison of the measurements of the late Universe, provided by galaxy surveys, and the early Universe, provided by CMB measurements, allows for powerful tests of the nature of cosmic acceleration and general relativity. The precision which photometric surveys are able to reach in the determination of cosmological parameters comes from the combination of different observables, mainly from weak lensing and clustering of galaxies, in the so-called \threextwo analysis, whose methodology is described in \citet{y3-generalmethods}.

In this work, we present the clustering measurements of the lens galaxy samples that enter in the DES Year 3 (Y3) \threextwo \citep{y3-3x2ptkp} and the \twoxtwo (\citealt{y3-2x2ptaltlensresults,y3-2x2ptbiasmodelling}; \citealt*{y3-2x2ptmagnification}; \citealt{y3-gglensing}, in combination with the shear field or galaxy-galaxy lensing) analyses. The cosmological information is extracted from the large-scale structure (LSS) measurements using the angular two-point correlation function that characterizes the spatial distribution of galaxies in tomographic photometric redshift bins. However, the measurement of the angular correlation function is affected by spatially varying survey properties that must be taken into account and corrected to extract the full cosmological power of DES. These systematic effects come from the observing conditions and translate into changes in the selection function across the observed footprint or with redshift.

As photometric surveys have become more extended in area, both the impact of these survey properties or observational effects, and the diminishing statistical errors, have spurred the development of a variety of techniques to correct for them in clustering measurements. Already in SDSS \citep{2002ApJ...579...48S, 2006ApJ...638..622M} and 2MASS \citep{2005ApJ...619..147M} cross-correlations  with different survey properties and masking were used to check for possible sources of systematic error, which were deemed to be insignificant given the statistical errors. \citet{Ross:2011} compared several methodologies (masking, cross-correlation correction and computing weights for the data) in SDSS-III. The cross-correlation correction method was applied to early DES data (DES-SV) in \citet{2016MNRAS.455.4301C}, and was studied by \citet{Elsner:2016} (there called ``template subtraction") who derived its characteristic bias. The application of weights have increasingly become a popular method, applied for instance in BOSS \citep{Ross:2017, 2020MNRAS.498.2354R}, eBOSS \citep{Laurent:2017}, DES-SV \citep[][ comparing with the cross-correlation method]{Kwan:2017}, DES Y1 data \citep{Elvin-Poole:2017xsf} and DESI targets \citep{2020MNRAS.496.2262K}. Rather than applying weights to the observed data, the inverse-weights can be applied to the random sample used for correlation function analyses, as shown in \citet{2015MNRAS.454.3121M} and applied to eBOSS data via a multilinear regression analysis in \citep{Bautista:2018,Icaza-Lizaola:2020}. These approaches have been refined in recent years as the importance of addressing these spatial systematics has grown \citep{Vakili:2020, Weaverdyck_2021, Wagoner_2021}, including the development of machine learning approaches using neural networks \cite{bossnn} or self-organizing maps \cite{Johnston:2021}. Some approaches have operated only at the level of the power spectrum, including mode projection methods (\citet{1992ApJ...398..169R} with examples of applications and further developments shown in \citet{2013MNRAS.435.1857L, 2014MNRAS.444....2L, Elsner:2016, Elsner:2017}). \citet{Weaverdyck_2021} reviewed several of the above techniques and showed how mode projection methods operating on the pseudo-power spectrum are related to multilinear regression methods, identifying residual biases in both approaches.
 
We present the methods we apply to DES-Y3 data in order to mitigate these effects, the full set of validation tests we perform, both on data and on simulations, and its final implementation on the data. These corrections enable robust measurements of the clustering amplitude of lens galaxies. The results of this analysis are used as the clustering input for the full \threextwo cosmological analysis in DES-Y3 \citep{y3-3x2ptkp}. 

This paper is organised as follows: in Section \ref{sec:theory} we describe the modeling of the galaxy clustering angular correlation function used throughout the Y3 analysis. In Section \ref{sec:samples}, we introduce the Y3 data and the galaxy samples derived from it. In Section \ref{sec:sp_maps}, we present the description of different observing conditions and their representation. In Section \ref{sec:methods}, we present the methodology, with special attention to the decontamination pipeline (subsections \ref{sec:pipeline} and \ref{sec:enet}). In Section \ref{sec:results}, we show the galaxy clustering results after applying the correction methods. This correction is validated in Section \ref{sec:weights_validation}. In Section \ref{sec:redmagic_discussion}, we discuss the post-unblinding findings about the amplitude of the angular correlation functions in terms of the considered survey properties. Finally, we present the conclusions in Section \ref{sec:conclusions}.

\section{Modelling}\label{sec:theory}
The observed projected galaxy density contrast $\delta_{\mathrm{obs}}^{i}(\hat{\mathbf n})$ of galaxies in tomography bin $i$ at position $\hat{\mathbf n}$ can be written as
\begin{equation}
\delta_{g,\mathrm{obs}}^{i}(\hat{\mathbf n}) = \underbrace{\int d\chi\, W^i_{\delta}\left(\chi \right)\delta_{g}^{(\rm 3D)}\left(\hat{\mathbf n} \chi, \chi\right)}_{\delta_{g,\mathrm{D}}^{i}(\hat{\mathbf n})} + \delta_{g,\mathrm{RSD}}^{i}(\hat{\mathbf n}) + \delta_{g,\mu}^{i}(\hat{\mathbf n})\,,
\end{equation}
with $\chi$ the comoving distance, $W_{\delta}^i = n_g^i(z)\, d z/d\chi$ the normalized selection function of galaxies in tomographic bin $i$. Here the first term is the line-of-sight projection of the three-dimensional galaxy density contrast, $\delta_{g}^{(\rm 3D)}$; the remaining terms are the contributions from linear redshift-space distortions (RSD) and magnification ($\mu$), which are described in \citet{y3-generalmethods}.

We model the galaxy density assuming a local, linear galaxy bias model \citep{FryGaztanaga:1993}, where the galaxy and matter density fluctuations are related by $\delta_g({\bf{x}})=b\delta_m(\bf{x})$, with density fluctuations  defined by $\delta \equiv (n({\bf{x}})-{\bar{n}})/{\bar{n}}$. We model the linear galaxy bias to be constant across each tomographic bin, denoted as $b^i$. The validity of these assumptions to the accuracy of the Y3 \threextwo analysis is demonstrated in \cite{y3-generalmethods}. 

The angular power spectrum consists of six different terms, corresponding to auto- and cross-power spectra of galaxy density, RSD, and magnification. At the accuracy requirements of the Y3 \threextwo analysis, the commonly-used Limber approximation is insufficient to evaluate these terms, and we adopt the non-Limber algorithm of \citet{Fang:2020}. For example, the exact expression for the density-density contribution to the angular clustering power spectrum is
 \begin{align}
  \nonumber   C_{\delta_{g,\rm D}\delta_{g,\rm D}}^{ij} (\ell)=&\frac{2}{\pi}\int d \chi_1\,W^i_{\delta}(\chi_1)\int d\chi_2\,W^j_{\delta}(\chi_2)\\
     &\int\frac{dk}{k}k^3 P_{gg}(k,\chi_1,\chi_2)j_\ell(k\chi_1)j_\ell(k\chi_2)\,,
 \label{eq:Cl-DD}
\end{align}
with $P_{gg}(k,z_1,z_2)$ the 3D galaxy power spectrum; the full expressions including magnification and redshift-space distortion are given in \cite{Fang:2020}. Schematically, the integrand in Eq.~\ref{eq:Cl-DD} is split into the contribution from non-linear evolution, for which un-equal time contributions are negligible so that the Limber approximation is sufficient, and the linear-evolution power spectrum, for which time evolution factorizes. \footnote{\url{https://github.com/xfangcosmo/FFTLog-and-beyond}} 

The angular correlation function is then given by
\begin{align}
    w^i(\theta) =& \sum_\ell \frac{2\ell+1}{4\pi}P_\ell(\cos\theta) C^{ii}_{\delta_{g,\mathrm{obs}}\delta_{g,\mathrm{obs}}}(\ell)~,
\end{align}
where $P_\ell$ are the Legendre polynomials. 

Throughout this paper, we use the \textsc{CosmoSIS} framework\footnote{\url{https://bitbucket.org/joezuntz/cosmosis}} \citep{Zuntz:2015} to compute correlation functions, and to infer cosmological parameters. The evolution of linear density fluctuations is obtained using the \textsc{CAMB} \citep{LewisBridle:2002} module\footnote{\url{http://camb.info}} and then converted to a non-linear matter power spectrum  $P_{NL}(k)$ using the updated {\Halofit} recipe \citep{Takahashi:2012}. 

We model (and marginalise over) photometric redshift bias uncertainties as an additive shift $\Delta z^{i}$ in the galaxy redshift distribution $n_{\rm g}^{i}(z)$ for each redshift bin $i$, 
\begin{equation}
n^{i}_{g}(z) \rightarrow n^{i}_{g}(z - \Delta z^{i}),
\label{eq:shift}
\end{equation}
and a stretch parameter to characterise the uncertainty on the width for some of the tomographic bins and samples,
\begin{equation}
n^{i}_{g}(z) \rightarrow n^{i}_{g}\left(\sigma_z^i[z-\langle z\rangle]+\langle z\rangle\right).
\label{eq:stretch}
\end{equation}

The priors on the $\Delta z^{i}$ and $\sigma z^{i}$ nuisance parameters are measured and calibrated directly using the angular cross-correlation between the DES sample and a spectroscopic sample, as described in \cite{y3-lenswz}. We use the same $\Delta z^{i}$ and $\sigma z^{i}$ as in the Y3 \threextwo analysis for all tests of robustness of the parameter constraints, as listed in Table \ref{tab:priors_info}.

\section{Data}\label{sec:samples}
The Dark Energy Survey collected imaging data with the Dark Energy Camera \citep[DECam;][]{Flaugher:2015} mounted on the Blanco 4m telescope at the Cerro Tololo Inter-American Observatory (CTIO) in Chile during six years, from 2013 to 2019. The observed sky area covers $\sim 5000 \deg^2$ in five broadband filters, $grizY$, covering near infrared and visible wavelengths. This work uses data from the the first three years (from August 2013 to February 2016), with approximately four overlapping exposures over the full wide-field area, reaching a limiting magnitude of $i\sim23.3$ for S/N = 10 point sources. The data were processed by the DES Data Management system \citep{Morganson:2018} and, after a complex reduction and vetting procedure, compiled into object catalogues. The catalogue used here amounts to nearly 400 million sources \citep[available publicly as Data Release 1\footnote{\url{https://des.ncsa.illinois.edu/releases/dr1};};][]{DESDR1:2018}. We calculate additional metadata in the form of quality flags, survey flags, survey property maps, object classifiers and photometric redshifts to build the \gold data set \citep*{y3-gold}. 

From this catalogue, we build the different galaxy samples for large-scale structure studies. For robustness, we decided to use two different types of lens galaxies, \maglim and \redmagic, which are used as lens samples for galaxy clustering and for combination with weak lensing for the \threextwo analysis. These two samples are described in the following subsections. \footnote{Moreover, from \gold \, we also define the \baosample , a galaxy sample especially defined for studies on the baryonic acoustic oscillation scales \citep{y3-baosample}, that is not used here, but undergoes an analogous treatment of its spatial systematics.}

\subsection{Y3 \maglim sample}

The main lens sample considered in this work, \maglim, is the result of the optimization carried out in \cite{y3-2x2maglimforecast}. The sample is designed to maximize the cosmological constraining power of the combined clustering and galaxy-galaxy lensing analysis (also known as 2$\times$2pt) keeping the selection criterion as simple as possible. The selection cuts, based on the table columns from \citet*{y3-gold}, are:

\begin{itemize}
    \item \texttt{flags\_foreground}=0 \& \texttt{flags\_footprint}=1 \& bitand(\texttt{flags\_badregions},2)=0 \& bitand(\texttt{flags\_gold},126)=0
    \item Star-Galaxy separation with \texttt{EXTENDED\_CLASS\_MASH\_SOF} = 3
    \item i < $4 \cdot z_{\rm phot} + 18$
    \item i > 17.5
\end{itemize}

The first cut is a quality flag to remove badly measured objects or objects with issues in the processing steps. It also removes problematic regions due to astrophysical foregrounds. The second cut removes stars from the galaxy sample. The faint magnitude cut in the $i$-band depends linearly on the photometric redshift, $z_{phot}$, and selects bright galaxies. The photometric redshift estimator used for this sample is the Directional Neighbourhood Fitting \citep[\dnf, ][]{deVicente:2016} algorithm \citep[see also][]{y3-2x2ptaltlensresults}, in particular its mean estimate using 80 nearest neighbors in colour and magnitude space, by performing a hyperplane fit. The brighter magnitude cut removes residual stellar contamination from binary stars and other bright objects. 

We split the sample into six tomographic lens bins, with bin edges $\mathrm{z_{phot}} = [0.20,0.40,0.55,0.70,0.85,0.95,1.05]$. These edges have been slightly modified with respect to \cite{y3-2x2maglimforecast} in  order  to  improve  the  photometric redshift calibration \citep{deVicente:2016}. We refer the reader to \cite{y3-2x2maglimforecast} for more details about the optimization of this sample and its comparison with \redmagic and other flux-limited samples. The main properties of the sample are summarized at the top panel of Table~\ref{tab:samples_info}.

\subsection{Y3 \redmagic sample}

The \redmagic\, algorithm selects luminous red galaxies (LRGs) according to the magnitude-colour-redshift relation of red sequence galaxy clusters, calibrated using an overlapping spectroscopic sample. This sample is defined by an input threshold luminosity $L_{\rm min}$ and constant co-moving density. The full \redmagic\, algorithm is described in \citet*{Rozo:2016}. \redmagic is the algorithm used for the fiducial clustering sample of the DES Y1 \threextwo cosmology analyses \citep{DESY1cosmo:2018, Elvin-Poole:2017xsf}, with some updates improving the redshift estimates and selection uniformity, besides the usage of new photometry from \gold.

We define the Y3 \redmagic sample in five tomographic lens bins, selected on the \redmagic redshift point estimate quantity \texttt{zredmagic}. The bin edges used are $\mathrm{z_{\redmagic}} = [0.15, 0.35, 0.50, 0.65, 0.80, 0.90]$. The first three bins use a luminosity threshold of $L_{\min} > 0.5 L_{*}$ and are known as the high density sample. The last two redshift bins use a luminosity threshold of $L_{\min} > 1.0 L_{*}$ and are known as the high luminosity sample.

The \redmagic selection also includes the following cuts on quantities from the \gold catalogue and \redmagic calibration,

\begin{itemize}
    \item Removed objects with \texttt{FLAGS\_GOLD} in 8|16|32|64
    \item Star galaxy separation with \texttt{EXTENDED\_CLASS\_MASH\_SOF} $\geq 2$ 
    \item Cut on the red-sequence goodness of fit $\chi^{2} < \chi^{2}_{\rm max}(z)$
\end{itemize}

The main properties of the sample are summarized in the bottom part of Table~\ref{tab:samples_info}. See \citet*{y3-gold} for further details on these quantities.

\begin{table}
    \centering
    \begin{tabular}{c|c|c|c|c}
        \multicolumn{5}{ | |c| | }{\maglim } \\ \hline
        \textbf{Redshift bin} & $N_g$ & $\langle n_g \rangle$ & $b^i$ & $\theta > [\mathrm{arcmin}]$ \\ \hline 
        $0.20 < z < 0.40$ & 2236462 & 0.150 & 1.5 & 33.88 \\ \hline 
        $0.40 < z < 0.55$ & 1599487 & 0.107 & 1.8 & 24.35 \\ \hline
        $0.55 < z < 0.70$ & 1627408 & 0.109 & 1.8 & 17.41 \\ \hline
        $0.70 < z < 0.85$ & 2175171 & 0.146 & 1.9 & 14.49 \\ \hline
        $0.85 < z < 0.95$ & 1583679 & 0.106 & 2.3 & 12.88 \\ \hline
        $0.95 < z < 1.05$ & 1494243 & 0.100 & 2.3 & 12.06 \\ \hline
        \hline
        \multicolumn{5}{ | |c| | }{\redmagic } \\ \hline
        \textbf{Redshift bin} & $N_g$ & $\langle n_g \rangle$ & $b^i$ & $\theta > [\mathrm{arcmin}]$ \\ \hline 
        $0.15 < z < 0.35$ & 330243 & 0.022 & 1.7 & 39.23 \\ \hline 
        $0.35 < z < 0.50$ & 571551 & 0.038 & 1.7 & 24.75 \\ \hline 
        $0.50 < z < 0.65$ & 872611 & 0.059 & 1.7 & 19.66 \\ \hline 
        $0.65 < z < 0.80$ & 442302 & 0.030 & 2.0 & 15.62 \\ \hline 
        $0.80 < z < 0.90$ & 377329 & 0.025 & 2.0 & 12.40 \\ \hline
    \end{tabular}
    \caption{\maglim (top table) and \redmagic (bottom table) characterisation parameters: number of galaxies, $N_g$, and number density, $\langle n_g \rangle$, blind galaxy bias, $b^i$ and scales excluded per redshift bin. The number densities are in units of $\mathrm{arcmin}^{-2}$ and the scales excluded correspond to 8 $\mathrm{Mpc}/h$ for both samples, as described in \protect\cite{y3-generalmethods}. The blind galaxy bias values correspond to the fiducial values that were assumed to create the log-normal mocks used in this analysis, not the best-fit values from 3$\times$2pt.}
    \label{tab:samples_info}
\end{table}

\subsection{Angular Mask}

The total sky area covered by the \gold catalogue footprint is $4946 \deg^2$. We then mask regions where astrophysical foregrounds (bright stars or large nearby galaxies) are present, or where there are known processing problems ("bad regions"), reducing the total area by $659.68 \deg^2$ \citep*{y3-gold}. The angular mask is defined as a \healpix\footnote{\url{https://healpix.sourceforge.io}} \citep{healpix} map of resolution $N_{\rm side}=4096$. Pixels with fractional coverage smaller than 80\% are removed. In addition, we require homogeneous depth across the footprint for both galaxy samples, removing too shallow or incomplete regions. As a summary, we use the following \gold and \redmagic specific map quantities to define the final common area:

\begin{itemize}
\item footprint = 1
\item foregrounds = 0
\item badregions $\leq 1$
\item fracdet > 0.8
\item depth $i$-band $\geq 22.2$
\item $z_{\rm MAX, high dens}$ $\geq 0.65$
\item $z_{\rm MAX, high lum}$  $\geq 0.95$
\end{itemize}
\noindent
where the depth for the $i$-band magnitude is obtained using the SOF photometry (detailed in \citealt*{y3-gold}) (as used in \maglim) and the conditions on ZMAX are inherited from the \redmagic redshift span. The final analysed sky area is $4143 \deg^2$.

\section{Survey properties}\label{sec:sp_maps}
\subsection{Survey property (SP) maps}
Through their impact on the galaxy selection function, survey properties can modify the observed galaxy density field. In order to correct these effects, we develop spatial templates for potential contaminants by creating \healpix \, sky maps of survey properties ("SP maps"), which we then use to characterize and remove contamination from the observed density fields \citep[see ][for the details of the original implementation of this mapping in DES]{Leistedt:2016}. Each pixel of a given SP map corresponds to a summary statistic that characterises the distribution of values of the measured quantity over multiple observations. Table \ref{tab:std_maps} summarizes the survey properties considered in this analysis along with the summary statistics used to produce the SP maps. As foreground sources of contamination we use a star map created with bright DES point sources, labeled \stellardens , and the interstellar extinction map from \cite{sfd98}, \sfd \, \footnote{We have verified that substituting the DES point sources map with the Gaia EDR3 star map (\cite{gaia_edr3}) and the \sfd \, map with the Planck 2013 thermal dust emission map (\cite{planck13}) has no significant impact on the results.}. More detailed information on the construction of these maps can be found in \citet*{y3-gold}. Hereafter we will use SP map to refer to survey property and foreground maps generically. 

\begin{table}
    \centering
    \begin{tabular}{c|c|c}
        \textbf{Quantity} & \textbf{Units} & \textbf{Statistics} \\ \hline
        \airmass & $\emptyset$ & WMEAN, MIN, MAX \\ \hline 
        \fwhm & arcsec & WMEAN, MIN, MAX \\ \hline 
        \fwhmfluxrad & arcsec & WMEAN, MIN, MAX \\ \hline
        \exptime & seconds & SUM \\ \hline 
        \teff & $\emptyset$ & WMEAN, MIN, MAX  \\ \hline
        \teffexptime & seconds & SUM \\ \hline
        \skybrite & electrons/CCD pixel & WMEAN \\ \hline
        \skyvar & (electron s/CCD pixel)$^2$ & WMEAN, MIN, MAX \\ \hline
        \skyvarsqrt & electrons/CCD pixel & WMEAN \\ \hline
        \skyvaruncert & electrons/ s $\cdot$ coadd pixel &  \\ \hline
        \sigmamagzero & mag & QSUM \\ \hline
        \fgcmgry & mag & WMEAN, MIN \\ \hline
        \maglimdepth & mag &  \\ \hline
        \sofdepth & mag &  \\ \hline
        \magautodepth & mag &  \\ \hline
        \stars & \# stars &  \\ \hline 
        \stellardens & stars/$\deg^2$ &  \\ \hline
        \sfd & mag &  \\ \hline
    \end{tabular}
    \caption{Survey properties used for the systematics mitigation effort of the DES Y3 Key Project, along with their physical units and the statistics used to generate SP maps from the stacking of images. As foreground sources of contamination we use a DES bright stars map and the dust extinction map from \protect\cite{sfd98}. We use both the raw number count of DES point sources, \stars, and the density, \stellardens. We use an SP map for each statistic in each photometric band in $\{g,r,i,z\}$ (with the exception of \stars, \stellardens and \sfd), resulting in 107 total SP maps.}
    \label{tab:std_maps}
\end{table}

\subsection{Reduced PCA map basis}\label{sec:pca_maps}
The Y1 analysis used 21 SP maps selected a priori. However,
a reduced set of SP maps is equivalent to setting a hard prior of no contamination from those SP maps that are unused, so we should be careful to not discard spatial templates that carry unique information about potential systematics \citep{Weaverdyck_2021}.
For Y3 we have initially increased the number of SP maps considered to 107.  
By expanding the library of SP maps used for cleaning, we relax the implicit priors and adopt a more data-driven approach to cleaning observational systematics from the clustering data. 

Many of the Y3 additional SP maps we use are alternative summary statistics for characterising the observed quantity, such as MIN and MAX instead of the weighted mean (WMEAN), which results in a high correlation between SP maps.
We therefore create an orthogonal set of SP maps by using the principal components of the pixel covariance matrix across all 107 SP maps (standardised to zero mean and unit variance) at $\nside = 4096$. This provides an orthornormal basis set of SP maps that can be ordered according to the total variance they capture in the space spanned by the 107 SP maps. We will refer to these principal component maps as PC maps to differentiate from SP maps in the standard (STD) basis, where each map represents a single survey property (e.g., \exptime). From this point forward, we will use ``SP'' map to more generically refer to maps that may be in either the PC or STD basis. We retain the first 50 PC maps, which account for $\sim98\%$ of the variance of the full 107 map basis. This allows us to capture the dominant features of the additional maps while reducing the risk of removing real LSS signal from overfitting. We test the impact of adjusting the number of PC maps used in Section \ref{sec:redmagic_discussion} and in App.~\ref{app:pc_cutoff}, finding that the full set of 107 maps results in galaxy weights that overcorrect and correlate significantly with LSS. The fiducial set of maps employed to decontaminate the data are these first 50 PC maps, although we have also run validation tests with the STD maps, as we explain in the next sections.

\section{Analysis Tools and Methodology}\label{sec:methods}
\subsection{Clustering Estimator}\label{cap:wtheta_estimator} 
The analysis of the galaxy clustering is performed by measuring the angular 2-point correlation function, $w(\theta)$, in photometric redshift bins. In this analysis we work with \healpix \, \citep{healpix} maps of the SPs and galaxy density from log-normal mock catalogues. The decontamination methods generate \healpix \, weight maps as well. Weights are actually obtained for each SP pixel, so we also work with pixelised versions of our galaxy samples, and use a pixel-based version of the Landy-Szalay estimator \citep{LS}, following the notation of \cite{2016MNRAS.455.4301C}:
\begin{equation}
\hat{w}(\theta) = 
\sum_{i=1}^{N_{pix}}\sum_{j=1}^{N_{pix}} \frac{(N_i-\Bar{N})\cdot(N_j-\Bar{N})}{\Bar{N}^2} \Theta_{i,\, j} \, ,
\end{equation} 
where $N_i$ is the galaxy number density in pixel $i$, $\Bar{N}$ is the mean galaxy number density over all pixels within the footprint and $\Theta_{i, \, j}$ is a top-hat function which is equal to $1$ when pixels $i$ and $j$ are separated by an angle $\theta$ within the bin size $\Delta \theta$. The fractional coverage of each pixel is taken into account in the calculation of $N_i$ and $\Bar{N}$. These correlation functions are calculated using \treecorr\footnote{\url{https://rmjarvis.github.io/TreeCorr}} \, \citep{treecorr}. We verify on the data that the difference between this pixel version of the estimator and that using random points is negligible for the angular scales we consider. 

\subsection{Log-normal mocks}\label{sec:mocks}
We rely on a set of log-normal mock realisations of the observed data to evaluate the significance of the correlation between data and SP maps following the methodology of \cite{Elvin-Poole:2017xsf} and \cite{Xavier:2016}. For each of our galaxy samples we create a set of $1000$ mocks that matches their mean galaxy number density and power spectrum. We generate full sky mock catalogues at a
\healpix \, resolution of $\nside = 512$, corresponding to $\sim 0.11$ degrees 
pixels. We then apply the DES-Y3 angular mask. This angular resolution is small enough to be used for the scales employed in the cosmology analysis. The usage of these mocks is covered in Section \ref{sec:pipeline}. We also create sets of contaminated log-normal mocks that we later use to validate our decontamination methods. These mocks incorporate the effect of SP maps observed on the data. Appendix \ref{app:app_mocks} contains more details about their creation and contamination. 

\subsection{Correction methods}\label{sec:correction}
The observed galaxy sample has contamination from observing conditions and foregrounds, which modify the selection function across the survey footprint. Our goal is to correct these effects in the lens galaxy samples. To do so, we create a set of weights to apply to the galaxy samples, constructed from a list of SP maps.  The weighted sample is then used for measurements of $w(\theta)$ and for combination with weak lensing measurements (\cite{y3-3x2ptkp}, \cite{y3-2x2ptaltlensresults}, \cite{y3-2x2ptbiasmodelling}, \cite{y3-2x2ptmagnification}). This approach has been successfully applied to the angular correlation function  of the DES Year 1 clustering measurements \citep{Elvin-Poole:2017xsf}, as well as in SDSS-III \citep[for example, in][]{Ross:2011, Ross:2017}, eBOSS \citep[][]{Laurent:2017, Bautista:2018, Icaza-Lizaola:2020, 2020MNRAS.498.2354R, 2021MNRAS.500.3254R} and in KiDS \citep{Vakili:2020}.

Most correction procedures can be interpreted as regression methods, with the true overdensity field corresponding to the residuals after regressing the observed density field against a set of SP maps. Adding SP maps is equivalent to adding additional explanatory variables to the regression, which increases the chance of overfitting. Such overfitting will reduce the magnitude of the inferred overdensity field (i.e. shrink the size of regression residuals), and thus overfitting will generically lead to a reduced clustering signal.

There are several approaches to address this. One can \textit{a priori} restrict the number of SP maps to reduce the level of false correction. This is equivalent to asserting that there is no contamination from the discarded SP maps, which risks biasing the data from unaccounted-for systematic effects. A second option is to clean with all of the SP maps and then debias the measured clustering based on an estimate of the expected level of false correction \citep[e.g. pseudo-$C_\ell$ mode projection, ][]{Elsner:2016, Elsner:2017, Alonso:2019}. This approach can be interpreted as a simultaneous ordinary least squares regression with a step to debias the power spectrum. Map-level weights that may enter in the analysis of other observables, such as galaxy-galaxy lensing, can be produced from this approach, but they will be overly-aggressive if the number of SP maps is large. \citet{Wagoner_2021} extend this approach by incorporating the pixel covariance and using Markov Chain Monte Carlo to include map-level error estimates, but this again becomes less feasible if the number of SP maps is too large. Finally, one can take an approach between these extremes, reducing the number of SP maps used for fitting, but doing so in a data-driven manner. We apply two different methods that take this third approach. They make different assumptions, but were both found to perform well in simulated tests in \cite{Weaverdyck_2021}. The SP maps we run these two methods on is our fiducial set of 50 PC maps that we introduced in Section \ref{sec:sp_maps}. In addition, we present a third method that we use to test linearity assumptions made by the other two. 

\subsubsection{Iterative Systematics Decontamination (\isd) }\label{sec:pipeline}
In this subsection, we describe the fiducial correction method that we use for DES Y3, called Iterative Systematics Decontamination (\isd). It is an extension of the
methodology applied in Y1 \citep{Elvin-Poole:2017xsf}. 

\isd is organised as a pipeline that corrects the PC map (or any generic SP map)
effects by means of an iterative process whose steps can  be summarized as i) identify the most significant PC  map, ii) obtain a weight map from it, iii) apply it to the data and iv) go back to i). The algorithm stops when there are no more maps with an effect larger than an a priori fixed threshold. Each step is described in more detail in the following lines. 

To begin with, we degrade each PC map to \nside = 512 and then we compute the relation between their values and $n_o / \langle n_o \rangle$ , where $n_o$ is the observed density of galaxies at a given part of the sky and $\langle n_o \rangle$ is the average density over the full footprint. In the following we refer to this as the 1D relation. To obtain the statistical significance of the observed correlations, we bin the 1D relation into ten equal sky areas for each PC map and 
estimate a covariance matrix for the 1D relation bin means of that PC map using the set of 1000 uncontaminated mocks described in Sec.\ref{sec:mocks}. Since the bins are defined as equal area, the statistical error associated with each bin is similar and no one region dominates the fit. We use this covariance matrix for determining the best-fit parameters of a function to approximate the 1D relation, as well as to assess its goodness-of-fit. 

We fit the 1D relation to a linear function of the PC map values
\begin{equation}
    \frac{n_{\rm o,i}}{\langle n_{\rm o}  \rangle} = m \cdot s_i + c \, ,
\end{equation}
by minimizing $\chi^2$, which we then denote $\chi^2_{\rm model}$. The index $i$ runs over the PC map bins. Similarly, we compute the goodness-of-fit for the case where $n_o / \langle n_o \rangle$ is a constant function $f(s) = 1$ labeled $\chi^2_{\rm null}$. Finding that $n_o / \langle n_o \rangle$ fits well to this constant function is equivalent to finding that this particular PC has no impact on the galaxy density field. To calculate both $\chi^2$ definitions, we make use of the ($10 \times 10$) covariance matrix obtained from the log-normal mocks. 

The degree of impact of a given PC map on the data is evaluated using
\begin{equation}
    \Delta \chi^2 = \chi^2_{\rm null} - \chi^2_{\rm model} \, . 
\end{equation}

To decide whether this impact is statistically significant or not, we run the exact same procedure described above on $1000$ log-normal mock realisations. In this way, we obtain the probability distribution of $\Delta \chi^2$. We define $\Delta \chi^2 (68)$ as the value below which are $68 \%$ of the $\Delta \chi^2$ values from the mocks. Then, we consider an SP map significant if 
\begin{equation}
    S_{1D} = \frac{\Delta \chi^2}{\Delta \chi^2 (68)} > T_{1D} \, ,
\end{equation}

\noindent where $T_{1D}$ is a significance threshold that is fixed beforehand. The square-root of this quotient is proportional to the significance in terms of $\sigma$. 

After identifying the most contaminating map, $s_i$, the next step is to obtain a weights map, $w_{s,i}$, to correct its impact. We compute this weights map as 
\begin{equation}\label{eqn:weights_definition}
    w_{s,i} = \frac{1}{F(s_i)} \, , 
\end{equation}

\noindent where $F(s_i)$ is a linear function of $s_i$ with which its 1D relation is fitted. In general, this function depends on the nature of the SP map, although the aim is to use functions as simple as possible to prevent overfitting. In the case of PC maps, we find no significant deviations from linearity in the 1D relations (see Appendix \ref{app:chi2_null_dist}).

\begin{figure}
    \includegraphics[width=\linewidth]{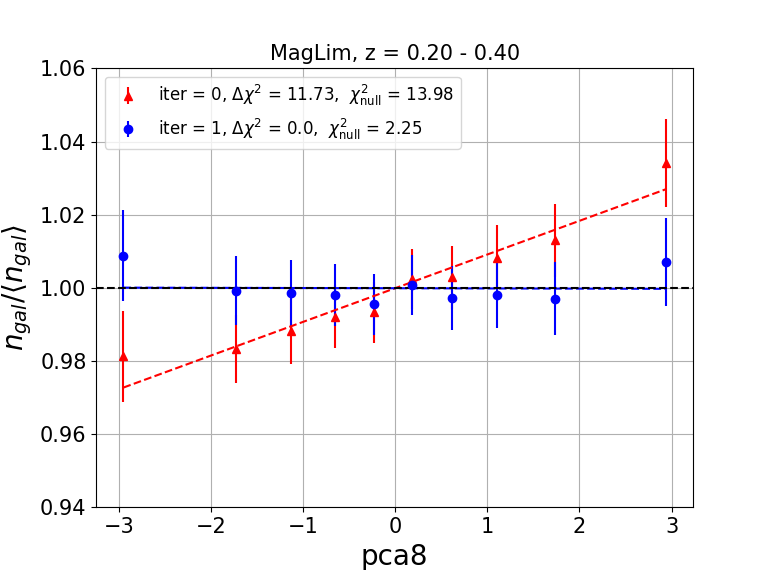}
    \caption{Example of how Iterative Systematics Decontamination (ISD) works. We illustrate this by showing the observed pixel number density (relative to the mean over the full footprint) as a function of a PC map pixel value, evaluated in ten equal area bins. We refer to this as 1D relation. The method identifies the PC map \pcmap{8} as the most significant one at iteration $0$ (i.e., no weights have been applied yet) at the first redshift bin of \maglim. The corresponding 1D relation is depicted by the red triangles and the red line corresponds to their best fit linear function. After correcting for the contaminating template with weights (given by equation \ref{eqn:weights_definition}) at iteration $1$, the impact of this PC map on the data is highly reduced. The blue points and their best fit linear function (blue line) show that the 1D relation is now compatible with no effect. }
    \label{fig:example_1ds}
\end{figure}

After obtaining the weight map, the pipeline normalises 
it to $\Bar{w}_s = 1$. Then, it is applied to the data, in
such a way that $N_{gal}^{p} \rightarrow N_{gal}^{p} \cdot w_s^p$, where $p$ is an index that runs over the footprint pixels at $\nside = 4096$. The process is repeated iteratively, identifying at each iteration the most significant PC map and correcting for it until all the PC maps have a significance lower than $T_{1D}$. At iteration $i$, the weights from iterations 1 to $i$ have been applied. Figure \ref{fig:example_1ds} shows the 1D relation of a given PC map that has been identified as a significant contaminant (dots) and after correcting for it (triangles). 

The weights associated to each significant PC map are incorporated multiplicatively to the total weight map, $w_T$, that is 
\begin{equation}\label{eq:totalweightmap}
    w_T = \prod_{i = 1}^f w_{s, \, i} \, ,
\end{equation}

\noindent where $i$ runs over the number of PC maps it is necessary to weight for. $w_T$ is then the total weight map that contains the information about the individual contaminants. These are the weights we apply to the data to mitigate the contamination. This total weight map is also normalised so its mean value over the full footprint is one. The pipeline runs this procedure for each redshift bin independently. 

\subsubsection{Elastic Net (\enet)} \label{sec:enet}
We also generated sets of weights using the Elastic Net (\enet) method described in \cite{Weaverdyck_2021} on the list of 50 PC maps. In this work, ENet has been used to perform robustness tests. Recall that the \isd \, method estimates contamination via a series of 1D regressions which are used to construct a total weight map via Eq.~\ref{eq:totalweightmap}. In contrast, \enet estimates the amplitude of contamination for all PC maps simultaneously, by maximizing the following log-posterior over $\alpha$:
\begin{equation}
\mathcal{P}(\alpha) \propto -\frac{1}{2N_{\rm pix}}||\delta_{\rm obs} - \mathbf{S}\alpha||_2^2 - \lambda_1 ||\alpha||_1 - \frac{\lambda_2}{2}||\alpha||_2^2, \label{eq:enetloss}
\end{equation}
where $\alpha_i$ is the contamination amplitude for PC map $s_i$, $\mathbf{S}$ is a matrix with the pixelated PC maps as columns\footnote{In practice, we standardise PC maps to have mean 0 and unit standard deviation before computing Eq.~(\ref{eq:enetloss}).}, and 
\begin{equation}
    \delta_{{\rm obs},j} = \frac{f_{\rm det, j}N_j}{\sum_j^{N_{\rm pix}}{(f_{\rm det, j}N_j)}/N_{\rm pix}} - 1,
\end{equation}
where $f_{\rm det, j}$ is the fraction of pixel $j$ that is not masked. The first term in equation~\ref{eq:enetloss} corresponds to the standard Gaussian likelihood that is maximized for an ordinary least squares regression. The regularizing terms act as components of a mixed, zero-centered prior on the elements of $\alpha$. The mixture consists of a Laplace and Gaussian distribution, with their precisions controlled by $\lambda_1$ and $\lambda_2$. The Laplace component is sharply peaked at zero, encouraging sparsity in the coefficients.
We determine the values of $\lambda_1$ and $\lambda_2$ by minimizing the mean squared error of the predictions on held-out portions of the footprint via 5-fold cross-validation. This allows the data to pick the precision and form of the prior based on predictive power.

We use the \texttt{scikit-learn} \citep{scikit-learn} implementation of \texttt{ElasticNetCV}, with a hyperparameter space of $\lambda_1/(\lambda_1+\lambda_2) \in \{0.1, 0.5, 0.9\}$ and 20 values of $(\lambda_1+\lambda_2)$ spanning four orders of magnitude (automatically determined from the input data). We degrade all maps to $\nside = 512$, and compute Eq.~(\ref{eq:enetloss}) using a \textit{training mask} that only includes pixels with $f_{\rm det} \geq 0.1$ (detection fraction from the \gold STD maps which is inherited by the PC maps). We performed many subsequent tests changing the definition of this training mask, with little observed impact on the final $w(\theta)$. Using \enet on the STD maps we also extended $\mathbf{S}$ to include quadratic terms of form $s_i^2$, and/or terms of form  $s_i s_{\rm stellardens}$, but these showed decreased predictive power on held-out samples, suggesting that the risk of overfitting from these additional maps dominates over additional contamination they identify. 

The total weight map is computed (still at $\nside = 512$) as
\begin{equation}
    w_T^{\enet} = \left[F_{\enet}(\mathbf{S})\right]^{-1} = (1 + \mathbf{S}\hat{\alpha})^{-1}.
\end{equation}

The \isd and \enet methods make different assumptions and take significantly different approaches to select important SP maps while minimizing the impact of overcorrection. \enet neglects the covariance of pixels, as well as the differing clustering properties of the SP maps, but it is less dependent on the basis of SP maps than is \isd. It avoids some of the difficulties the \isd method has when SP maps are highly correlated or contamination is distributed weakly across a combination of many maps, and hence missed by 1D marginal projections.  We therefore expect the \enet method to be a useful robustness test of the fiducial \isd method, and it is also used to estimate the systematic contribution to the $w(\theta)$ covariance (see Sec.~\ref{sec:results}).

\subsubsection{Neural net weights (NN-weights)}\label{sec:nnweights}
To evaluate the robustness of the assumptions made and codes used in producing galaxy-density weights, we created a third alternative process with different choices and independent code---in particular, abandoning the assumption that the mean galaxy density is a linear or polynomial function of all SP maps. The basic principle remains the same, namely that a function $w(\mathbf{s})$ of the vector $\mathbf{s}$ of SP values is found which maximizes the uniformity of the observed catalogue.  In this case, however, the function is realized by a neural network (NN), in a manner very similar to that of \citet{bossnn}.

In contrast to \isd and \enet, we apply this method on the STD basis of maps. In addition, two important changes to the weighting procedure were made to avoid having the NN overtrain, in the sense of absorbing true cosmological density fluctuations into the observational density factor $w.$  First, the input STD maps were limited to those which should in principle fully describe the characteristics of the coadd images: the \fwhm , \skyvaruncert , \exptime \, and \fgcmgry \, exposure-averaged values for each of the $griz$ bands, the \sfd \, extinction estimate, and a \textit{gaia\_density} estimate of local stellar density constructed from Gaia EDR3 \citep{gaia_edr3}.  We confirm that weights constructed with these STD maps eliminate any correlation of galaxy density on \airmass \, or \textit{depth}, and additionally find that \fgcmgry \, has no significant effect, so it is dropped, leaving 14 STD maps. The second major change to avoid overtraining is to institute $N$-fold cross-validation: the footprint is divided into healpixels at $\nside = 16$, which are randomly divided into $N$ distinct ``folds.''  The weights for each fold are determined by training the NN on the other $N-1$ folds, halting the training when the loss function for the target fold stops improving.  We use $N=3.$

The weights are created on a healpixelization at $\nside = 4096$. With $n_i, f_i,$ and $w_i$ being the galaxy counts, useful-area fraction, and weight estimate for each healpixel, the NN is trained to minimize the binary cross-entropy
\begin{equation}
  S \equiv \sum_{n_i>0} \log \bar n f_i w_i +  \sum_{n_i=0} \log\left(1-\bar n f_i w_i\right).
\end{equation}
In a further departure from the standard weighting scheme, we take the input vector $\mathbf{s}$ to be the logarithm of each input STD map (except for \sfd , which is already a logarithmic quantity), then linearly rescale each dimension to have its 1--99 percentile range span $(0,1)$.  We mask the $<1\%$ of survey area for which any such rescaled SP has $s_i$ outside the range $(-0.5,1.5),$ knowing that the NN will fail to train properly on rare values of STD maps.

Using the Keras software\footnote{\url{https://keras.io}}, we define the weight function for a given galaxy bin as
\begin{equation}
  \log w(\mathbf{s}) = \boldsymbol{\alpha}\cdot \mathbf{s} + NN(\mathbf{s}),
\end{equation}
where $\boldsymbol{\alpha}$ defines a nominal power-law relationship between the STD maps and the expected galaxy density, and $NN$ is a three-layer perceptron describing deviations from pure power-law behavior. The training of all folds for all redshift bins can be done overnight on a single compute node.

\section{Results}\label{sec:results}
\isd returns a list of maps with significant impact on galaxy clustering and that we need to weight for in each redshift bin of the samples. We studied the impact of observing conditions at three different significance threshold values, $T_{1D} = 2, \, 4, \, 9$. Increasing this threshold is equivalent to relaxing the strictness of the decontamination, decreasing the number of significant SP maps. After testing for over and undercorrection on mocks, the fiducial choice of significance threshold is $T_{1D} = 2$ (see Sections \ref{sec:weights_validation} and \ref{sec:redmagic_discussion} for more details). 

\begin{figure*}
    \centering
    \includegraphics[width=\linewidth]{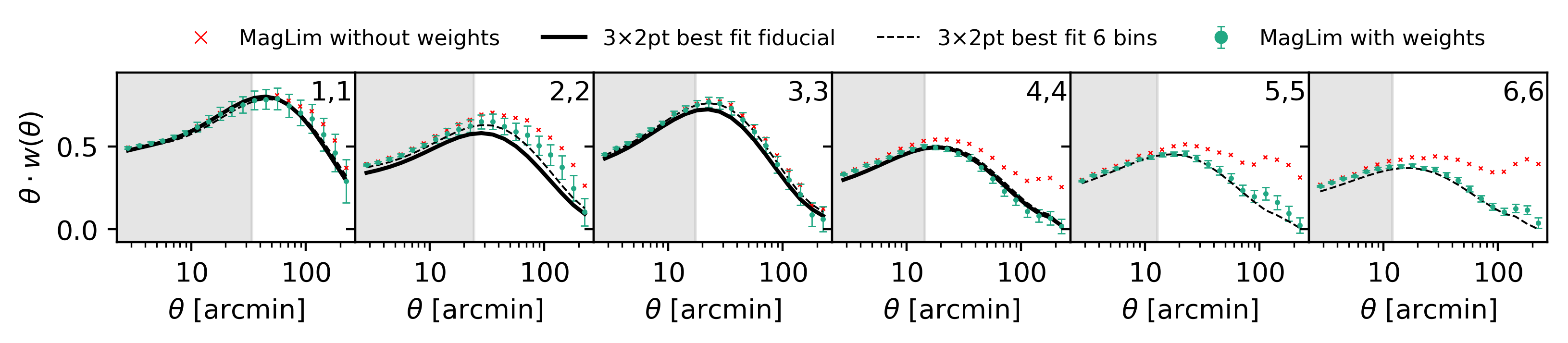}
    \includegraphics[width=0.95\linewidth]{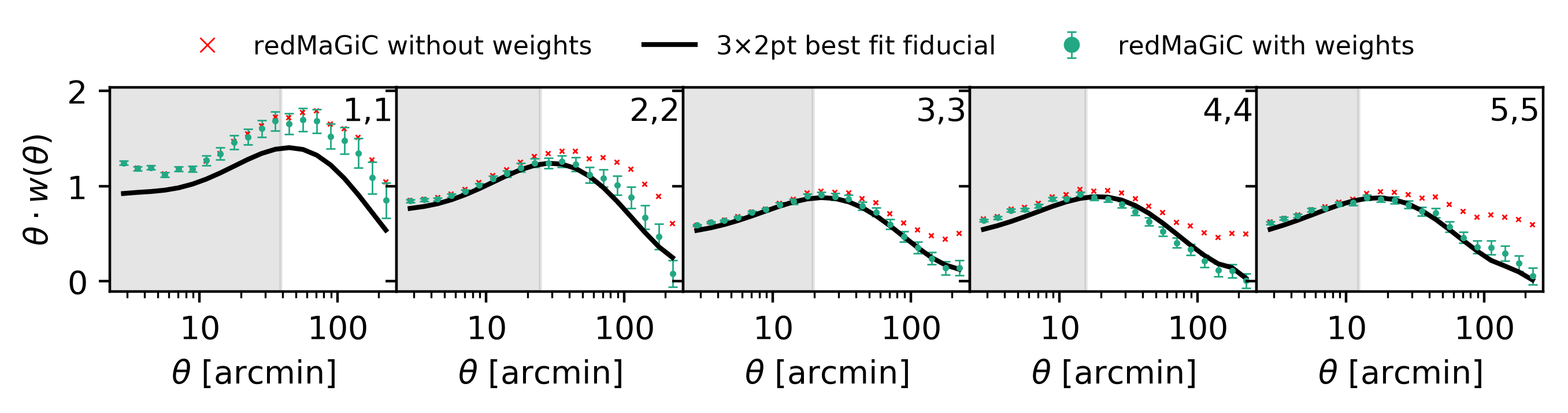}
    \caption{DES Y3 galaxy clustering results for \maglim (top panel) and \redmagic (bottom panel). The green points correspond to the angular correlation function of the \isd -PC<50 weighted data, while the red points correspond to the uncorrected data. The solid black line shows the best-fit theory prediction from the DES Y3 \threextwo $\Lambda$CDM results of each sample \citep{y3-3x2ptkp}. Note that for \maglim we also show the best-fit from the analysis including all six redshift bins (dashed black line), although the fiducial \threextwo cosmology results from this sample only include its first four bins. The shaded regions correspond to the scales that are excluded for cosmological constraints. }
    \label{fig:wtheta_data}
\end{figure*}

We find that, in general, both samples show a similar trend and they are more impacted by observing conditions at higher redshift. Generally, more SP maps are significant for the \maglim sample than for \redmagic . The measured angular 2pt correlation functions on the weighted samples can be seen in Figure \ref{fig:wtheta_data}. The S/N \footnote{The signal-to-noise is defined as $S/N \equiv \frac{w_{\rm data}(\theta) C^{-1} w_{\rm model}(\theta) }{\sqrt{w_{\rm model}(\theta) C^{-1} w_{\rm model}(\theta)}}$, where $C$ is the $w(\theta)$ part of the covariance matrix and $w_{\rm model}(\theta)$ is the best fit model from \threextwo .} of this detection is $\sim 63$ for both samples (using only the first four bins of \maglim). The data have been corrected for systematic contamination by applying the \isd -PC<50 weights. After the correction they are in good agreement (green points) with the best fit cosmology from 3$\times$2pt. The deviation in the first redshift bin for \redmagic is known to come from an inconsistency between clustering results and galaxy-galaxy lensing in this sample. We defer the discussion of this important result from the point of view of observational systematics to Section \ref{sec:redmagic_discussion}. We note also that for \maglim we depict two best fit correlation functions: the best fit model from 3$\times$2pt analysis using its six redshift bins (dashed black lines) and excluding its last two bins (solid black lines). The DES fiducial constraints are obtained without the last two bins, as explained in \cite{y3-2x2ptaltlensresults}. The shaded regions in this figure depict the scales excluded (see Table \ref{tab:samples_info}) from our data vectors. These regions are not used to obtain constraints on cosmological parameters. The uncorrected $w(\theta)$ are shown as red crosses. We note that the impact of systematic corrections is easily larger than the statistical uncertainty in the measurements, and are therefore necessary for unbiased cosmological inference, as we will illustrate below. These corrections are more important at higher redshift bins in both galaxy samples. For a comparison of this correction with respect to DES Y1 galaxy clustering, see \cite{Elvin-Poole:2017xsf}. 

\begin{figure}
    \centering
    \includegraphics[width=\linewidth]{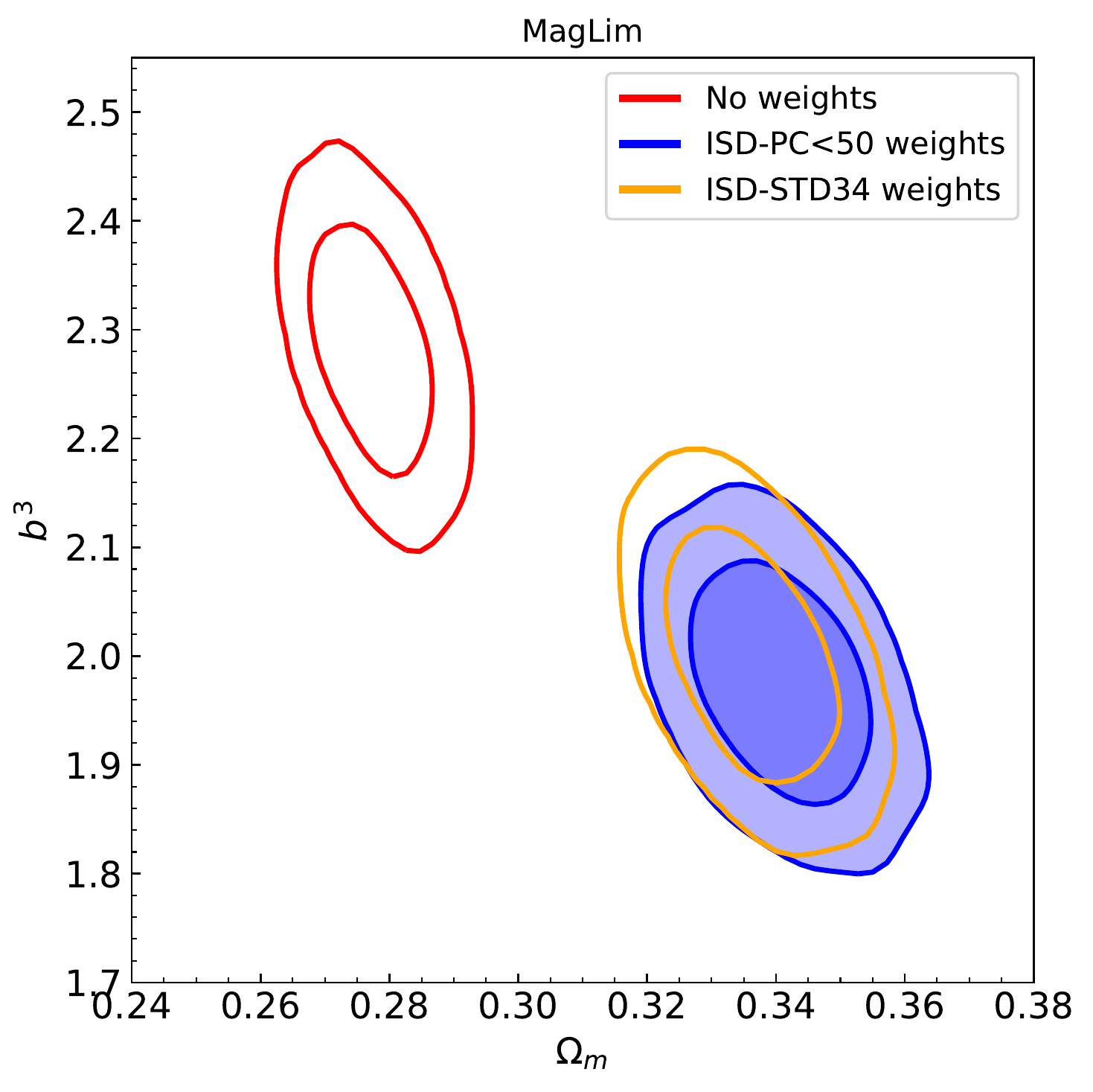}
    \includegraphics[width=\linewidth]{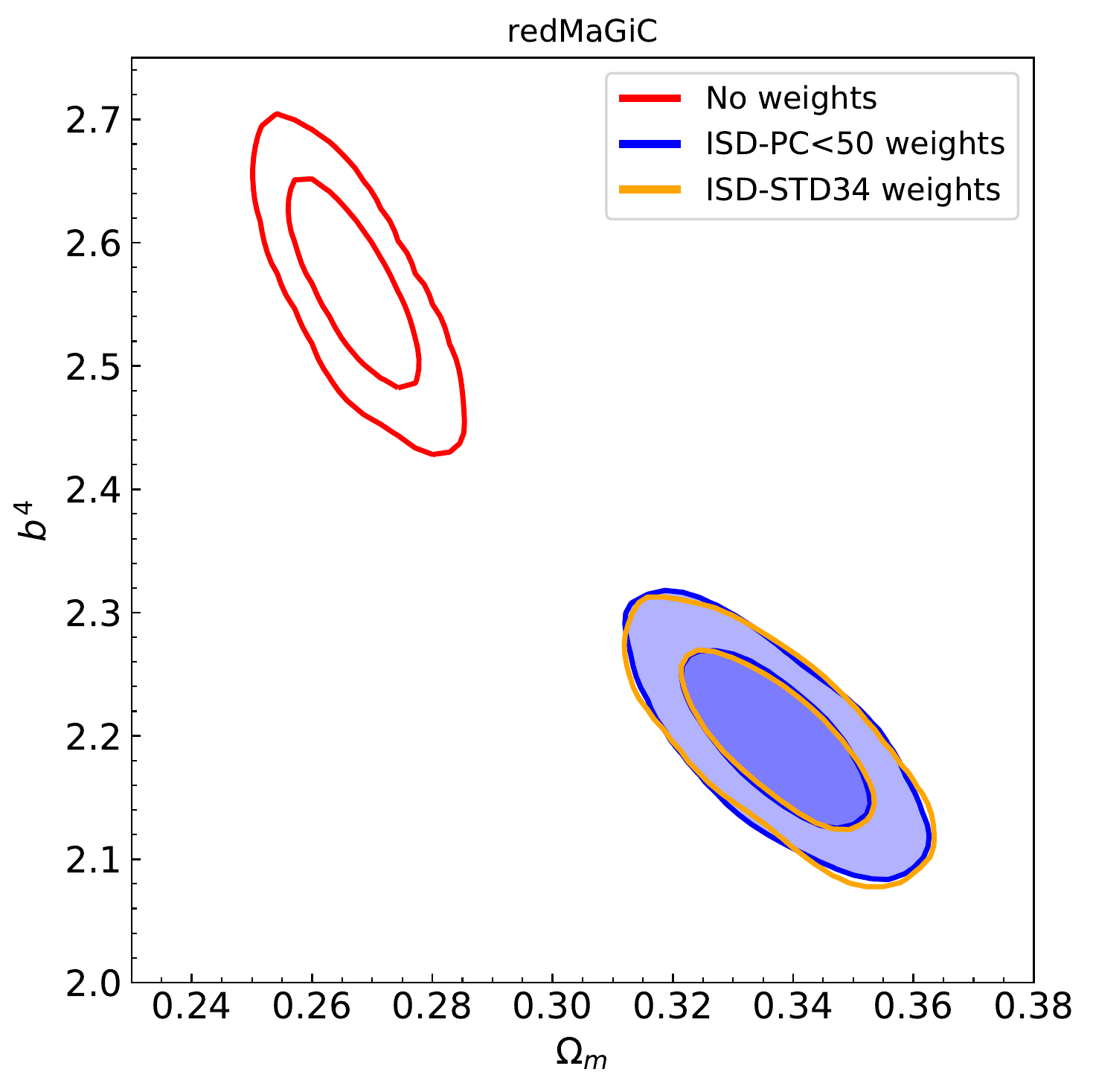}
    \caption{Constraints on $\Omega_m$ and galaxy bias before and after applying our weighting methodology to the data for \maglim (top panel) and \redmagic (bottom panel). We focus on the redshift bins where the difference in the mean posteriors of these parameters from contaminated (red contours) and decontaminated (filled blue contours) data is the greatest. The absence of correction strongly biases our estimations. We also show constraints for \isd -STD34 weighted data (orange contours). We obtain similar behaviours for the rest of the redshift bins of both samples. }
    \label{fig:contours_data}
\end{figure}

In Figure \ref{fig:contours_data}, we explicitly demonstrate the importance of our systematics correction by placing constraints on $\Omega_m$ and the clustering biases $b^i$ from the galaxy clustering correlation function alone. We do this by fitting the theory model presented in Section \ref{sec:theory} to the data using \cosmosis \, and the \texttt{PolyChord} sampling software \citep{Polychord1,Polychord2}. The covariance that we employ is given by \cosmolike \, \citep{Krause:2016jvl} and it includes the systematic contributions that we introduce in Section \ref{sec:cov_sys_terms}. We again marginalise over shifts in the photometric redshift distributions and over their widths. These nuisance parameters are sensitive to the clustering amplitude. For \redmagic the rest of the cosmological parameters are fixed to the DES Y3 fiducial best fit cosmology and for \maglim these are fixed to the best fit cosmology using the six redshift bins. For this reason, this constraint on $\Omega_m$ should not be taken as a true constraint, but this illustrates how the changes in the measured $w(\theta)$ can impact cosmology constraints. The priors for these cosmological and nuisance parameters are given in Table \ref{tab:priors_info}. We obtain these contours for the unweighted and \isd-weighted data. As evidence of robustness of our choice of SP maps, we also show contours for another configuration of \isd (\isd -STD34), where only 34 STD maps are considered (see Section \ref{sec:stdmapbasis} and Appendix B of \cite{y3-baosample} for more details on this selection of SP maps). 
We see that failure to apply our systematic corrections biases the inferred bias values as well as the recovered matter density relative to our fiducial choice. The corrections for the two \isd configurations are equivalent within the statistical uncertainty. In Figure \ref{fig:contours_data}, we focus on the redshift bins with the most prominent difference in the mean of the posteriors from uncorrected (red contours) and corrected data (blue contours). We find $4.10 \sigma$ and $6.96 \sigma$ differences in $b^3$ and $\Omega_m$, respectively, for \maglim. In the case of \redmagic, we find $7.69 \sigma$ and $6.79 \sigma$ differences in $b^4$ and $\Omega_m$. The effect of not correcting is to shift the contours towards higher galaxy biases and lower $\Omega_m$ values. This highlights the importance of correcting systematic effects. 


\begin{table}
    \centering
    \begin{tabular}{c|c|c} \hline\hline
        \multicolumn{3}{ |c| }{\maglim } \\ \hline
        \textbf{Redshift bin} & $\Delta z$ & $\sigma_z$ \\ \hline 
        $0.20 < z < 0.40$ &   (-0.009,0.007) &  (0.975,0.062) \\ \hline 
        $0.40 < z < 0.55$ &   (-0.035,0.011) &  (1.306,0.093) \\ \hline
        $0.55 < z < 0.70$ &   (-0.005,0.006) &  (0.87,0.054) \\ \hline
        $0.70 < z < 0.85$ &   (-0.007,0.006) &  (0.918,0.051) \\ \hline
        $0.85 < z < 0.95$ &   (0.002, 0.007) &  (1.08,0.067) \\ \hline
        $0.95 < z < 1.05$ &   (0.002, 0.008) &  (0.845,0.073) \\ \hline
        \hline
        \multicolumn{3}{| c | }{\redmagic } \\ \hline
        \textbf{Redshift bin} & $\Delta z$ &  $\sigma_z$ \\ \hline 
        $0.15 < z < 0.35$ &   (0.006,0.004) &  fixed to 1 \\ \hline 
        $0.35 < z < 0.50$ &   (0.001,0.003) &  fixed to 1 \\ \hline 
        $0.50 < z < 0.65$ &   (0.006,0.004) &  fixed to 1 \\ \hline 
        $0.65 < z < 0.80$ &   (-0.002,0.005) &  fixed to 1 \\ \hline 
        $0.80 < z < 0.90$ &   (-0.007,0.010)  & (1.23,0.054) \\ \hline \hline
        \multicolumn{3}{ | c| }{Both samples } \\ \hline
        & $\Omega_M$ & $b^i$  \\ \hline 
        All redshifts & [0.1,0.9] &  [0.8,3.0] \\ \hline 
    \end{tabular}
    \caption{List of prior values used to constrain $\Omega_M$ and the sample galaxy biases $b^i$ per redshift bin. The other cosmological parameters have been fixed to the fit values in the \threextwo analysis as described in the text. Square brackets denote a flat prior, while parentheses denote a Gaussian prior of the form $\mathcal{N}(\mu,\sigma)$. The shift $\Delta z$ and stretch $\sigma_z$ parameters are defined in Eqs.~(\ref{eq:shift},\ref{eq:stretch}). In some cases the latter is not marginalised over (fixed). The redshift priors were determined in \protect\cite{y3-lenswz}. }
    \label{tab:priors_info}
\end{table}

\section{Weights validation}\label{sec:weights_validation}
We validate our methodology on simulated catalogues to ensure that no biases are induced. We use unaltered log-normal mocks and also mocks that are artificially contaminated by our SP maps (see Appendix \ref{app:app_mocks} for details on how we apply this contamination). We contaminate these mocks by applying the inverse of the weights determined from the data using \enet on the full list of 107 STD maps. Decontamination, however, is performed using weights determined by \isd -PC<50. This procedure adds an additional layer of protection: if we contaminate mocks with the weights from one method and decontaminate by the same method, the test is only checking sensitivity to forms of contamination to which we \textit{a priori} know the method is sensitive to. Generating an equally plausible realization of contamination from an alternative method adds the benefit of potentially revealing blind spots in the method that is being validated. 

We calculate $\Bar{w}_{\rm dec}(\theta)$ and $\Bar{w}_{\rm unc}(\theta)$ as the mean correlation function of 400 decontaminated and 400 uncontaminated mocks, respectively. Since the log-normal mocks are generated at $\nside = 512$, which corresponds to separation angles of $\sim 6.9$ arcmin between pixels, we compute the correlation functions at the 14 fiducial angular scales that are larger than this limit.  Then we estimate the impact of the different biases (see next two Sections) on $w(\theta)$ by means of the true mean in uncontaminated mocks, $\Bar{w}_{\rm unc}(\theta)$:
\begin{equation}
    \chi^2 = (\Bar{w}_{\rm dec}(\theta)-\Bar{w}_{\rm unc}(\theta))^{\top} \cdot \, C^{-1} \cdot \, (\Bar{w}_{\rm dec}(\theta)-\Bar{w}_{\rm unc}(\theta)) \, . 
\end{equation}
The covariance matrix, $C$, is the galaxy clustering part of the analytical covariance given by \cosmolike, and it is also used for the clustering part of the \threextwo cosmological analysis. If we find that any bias causes a change in the joint fit to all redshift bins according to the definition above, equivalent to $\chi^2>3$,  then we marginalise over this bias in our final analysis. This threshold was chosen such that the impact on $\chi^2$ would be a small compared to the expected width of the $\chi^2$ distribution of the $w(\theta)$ data vector. As we detail in Section \ref{sec:cov_sys_terms}, we marginalise over biases by modifying the covariance matrix to account for these sources of systematic uncertainty. The fiducial covariance matrix for DES Y3 \threextwo analysis includes these systematic terms.

\subsection{False correction test}\label{sec:false_correction_bias}
Since we consider a large number of SP maps in this analysis, chance correlations between the data and some of these maps could arise, even after reducing our number of SP maps. This is more important when using a strict significance threshold. These purely random correlations could cause overcorrections, therefore biasing the measured value of $w(\theta)$ and the inferred cosmological parameters. To characterise this effect, we run \isd with $T_{1D} = 2$ on a set of 400 uncontaminated mocks and then we obtain their correlation functions, $w^{T_{1D}}_{w, \, \rm unc, \, \textit{i}}$. The false correction bias is defined as \\
\begin{equation}
w^{T_{1D}}_{\rm f. \, c. \, bias}(\theta) = \frac{1}{400} \, \left( \sum_{i=1} ^N w^{T_{1D}}_{w, \, \rm unc, \, \textit{i}} (\theta) \, - \, \sum_{j=1} ^N  w_{\rm unc, \, \textit{j}} (\theta) \right) \, ,
\end{equation}\\
where $w_{\rm unc, \, \textit{j}}$ are the correlation functions measured on the unaltered uncontaminated mocks. 

\begin{figure}
    \centering
    \includegraphics[width=\linewidth]{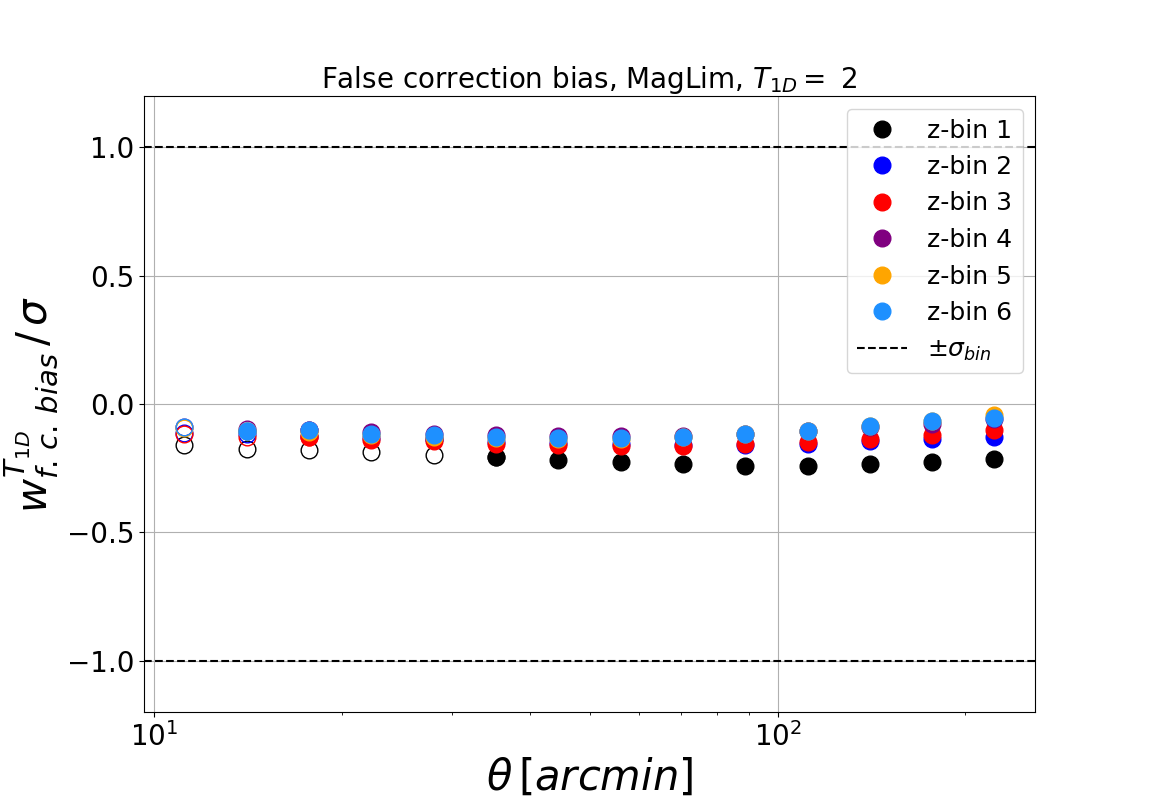}
    \includegraphics[width=\linewidth]{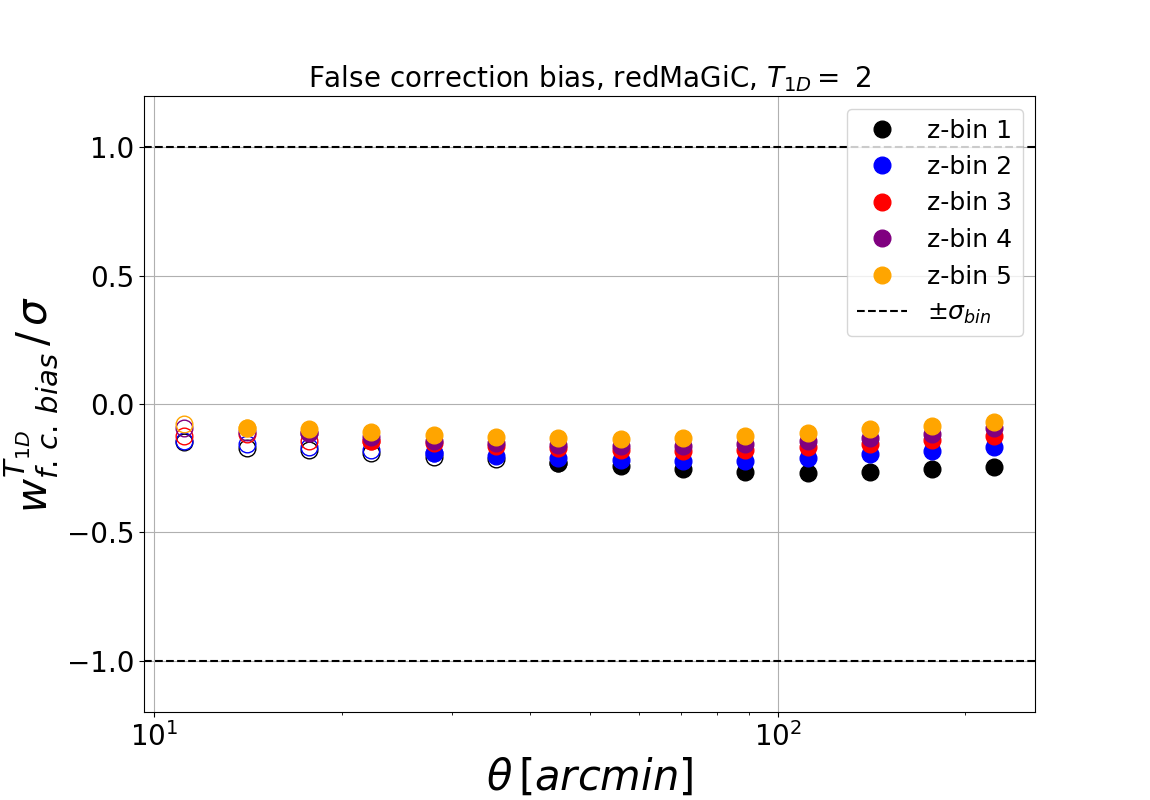}
    \caption{False correction bias, $w^{T_{1D}}_{\rm f. \, c. \, bias}(\theta)$, for \maglim(top panel) and \redmagic(bottom panel) relative to the $w(\theta)$ error from the unaltered \cosmolike \, covariance diagonal elements. Negative values are indicative of overcorrection. Both samples show negligible levels of overcorrection, weak dependence with the angular scale and at most $\sim 20\%$ of the statistical error. The values depicted here have been calculated with significance threshold $T_{1D} = 2$. Empty dots correspond to the angular scales not considered for each redshift bin of the samples. }
    \label{fig:false_bias}
\end{figure}

\begin{figure}
    \centering
    \includegraphics[width=\linewidth]{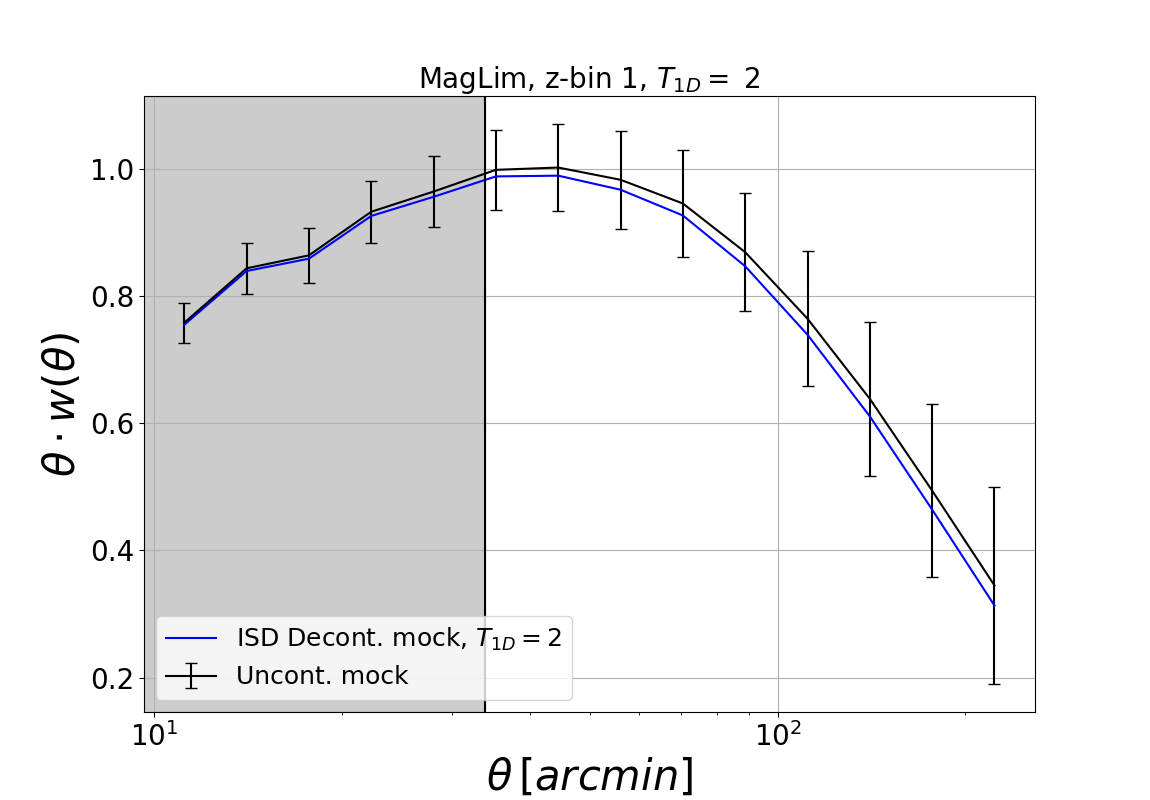}
    \includegraphics[width=\linewidth]{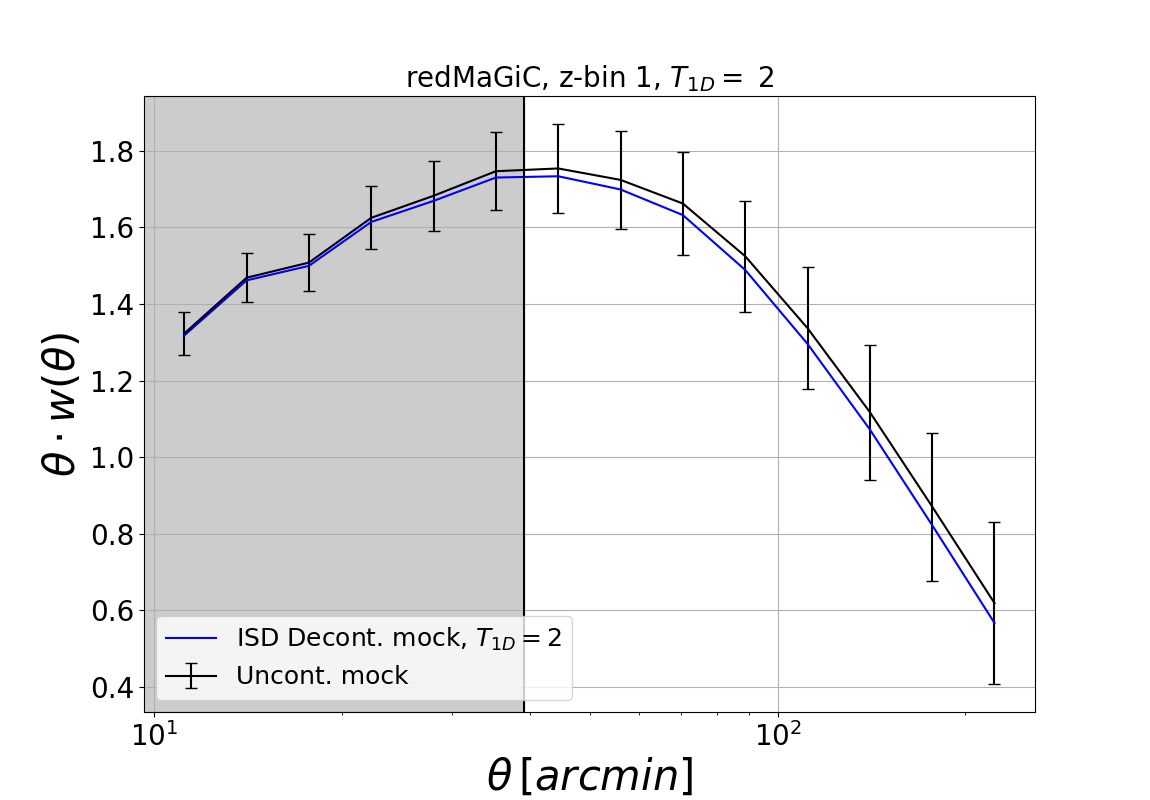}
    \caption{Mean angular correlation function, $w(\theta)$, from raw uncontaminated log-normal mocks (black lines) and from decontaminated uncontaminated mocks (blue lines) for \maglim (top panel) and for \redmagic (bottom panel) at their lowest redshift bins. Shaded region correspond to the scales excluded at this redshift. In this redshift bin there is $\sim 20\%$ of false correction with respect to the statistical error due to chance correlations between PC maps and mocks. The error bars correspond to the diagonal of the covariance matrix with systematic terms added. }
    \label{fig:wtheta_false_corr_bias}
\end{figure}

In general, the effect of removing the systematic effects is to diminish the amplitude of $w(\theta)$. Thus, a negative value of this estimator indicates overcorrection. In Figure \ref{fig:false_bias} we show the results of $w^{T_{1D}}_{\rm f. \, c. \, bias}(\theta) / \sigma$ for $T_{1D} = 2$, where $\sigma$ is the diagonal of the unmodified covariance matrix. We find a very marginal indication of overcorrection, always well below the statistical error. We also note that this ratio has small angular dependence, as can be seen in Figure \ref{fig:wtheta_false_corr_bias} which compares the mean true $w(\theta)$ (black line) with the mean of the decontaminated correlation functions (blue line).  Therefore, we do not consider any contribution from the false correction bias to the final covariance matrix. The small impact of this effect on the cosmological parameters is highlighted in Section \ref{sec:impact_on_parameters}. Nevertheless, we note that the error bars shown in Figure \ref{fig:wtheta_false_corr_bias} correspond to the diagonal of the covariance matrix which has been modified to account for systematic uncertainties, as it is explained in Section \ref{sec:cov_sys_terms}. 

\subsection{Residual systematic test}\label{sec:residual_systematic_bias}
Here we demonstrate that \isd effectively recovers the true correlation function from a contaminated sample. We can then verify if our approach (with $T_{1D} = 2$) meets the requirements for the Y3 cosmology analysis or whether it is necessary to account for any bias due to uncorrected contamination. 

We define the residual systematic bias as   
\begin{equation}
w^{T_{1D}}_{\rm r. \, s. \, bias}(\theta) = \frac{1}{400} \, \left( \sum_{i=1} ^N w^{T_{1D}}_{\rm dec, \, \textit{i}} (\theta) \, - \, \sum_{j=1} ^N  w_{\rm unc, \, \textit{j}} (\theta) \right) \, ,
\end{equation}\\
where the $w^{T_{1D}}_{\rm dec, \, \textit{i}}$ are the correlation functions measured on mocks that have had systematic contamination added and then have been decontaminated using \isd. 

Because we are interested in the level of \textit{residual} systematics that are insufficiently captured by the weighting method, we use the alternative method \enet with all 107 maps in the standard basis to generate an aggressive level of contamination. We observe that both \isd -PC107 and \enet -STD107 significantly overcorrect at the lowest redshift bins of both galaxy samples (see Section \ref{sec:redmagic_discussion}), so when using the corresponding weights to contaminate the mocks we are introducing excessive contamination. Therefore, we expect some degree of undercorrection when later running \isd with a sub-set of PC maps such as with \isd -PC<50. Furthermore, by using \enet to estimate the contamination instead of \isd, the contaminated mocks will include possible contamination modes to which \enet is sensitive but to which \isd may not be.  

\begin{figure}
    \centering
    \includegraphics[width=\linewidth]{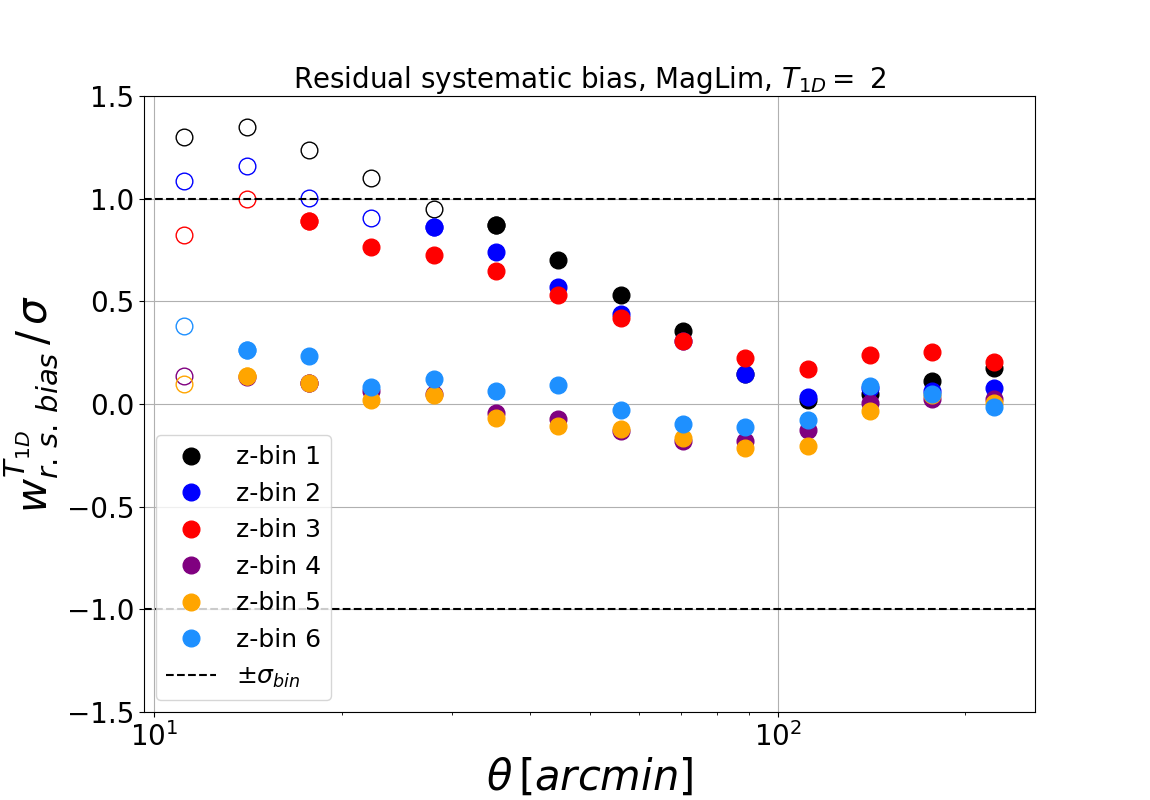}
    \includegraphics[width=\linewidth]{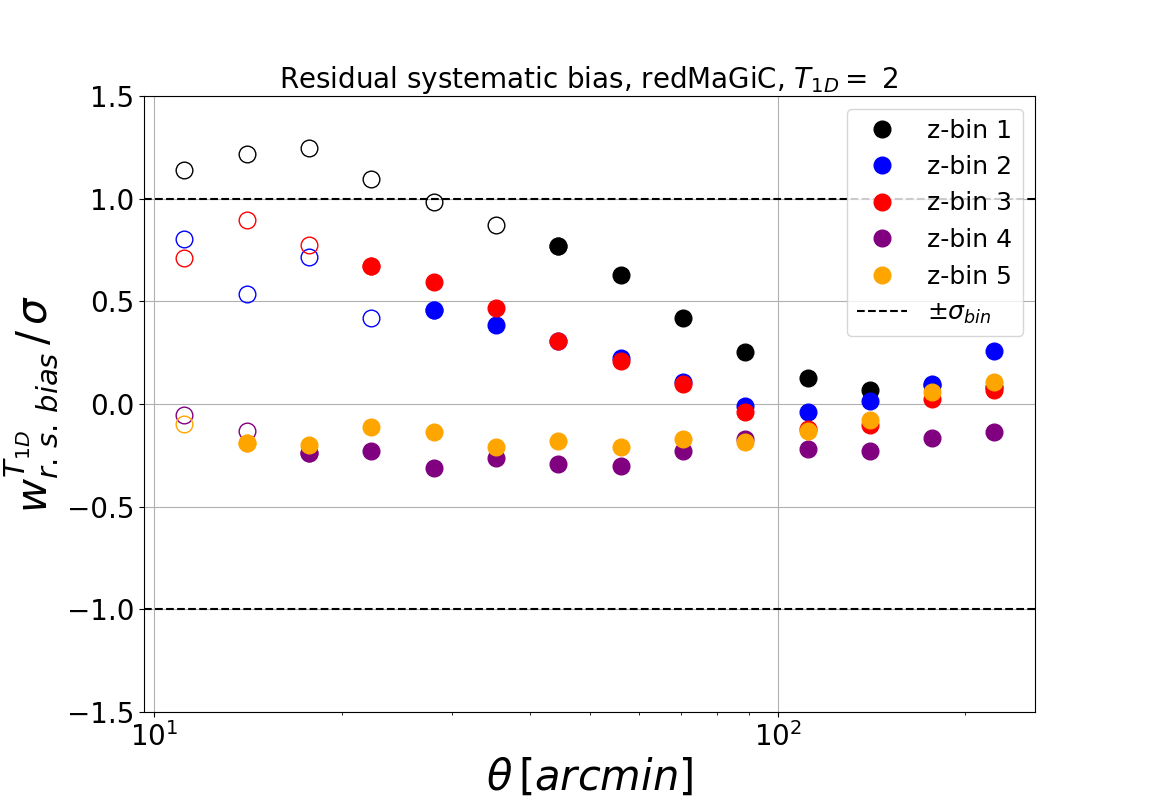}
    \caption{Residual systematic bias, $w^{T_{1D}}_{\rm r. \, s. \, bias}(\theta)$, for \maglim (top panel) and \redmagic (bottom panel) relative to the $w(\theta)$ error from the unaltered \cosmolike \, covariance diagonal. The empty dots represent the scales excluded at each bin. Both samples show similar trends: the highest redshift bins present lower biases, while the lowest ones show important levels of undercorrection at the smallest scales. On the other hand, the largest scales are recovered nearly unbiased. Since the $\chi^2$ of the total residual bias in all bins is higher than 3, we add a systematic term to the covariance matrix to marginalise over this effect. }
    \label{fig:resi_bias}
\end{figure}

\begin{figure}
    \centering
    \includegraphics[width=\linewidth]{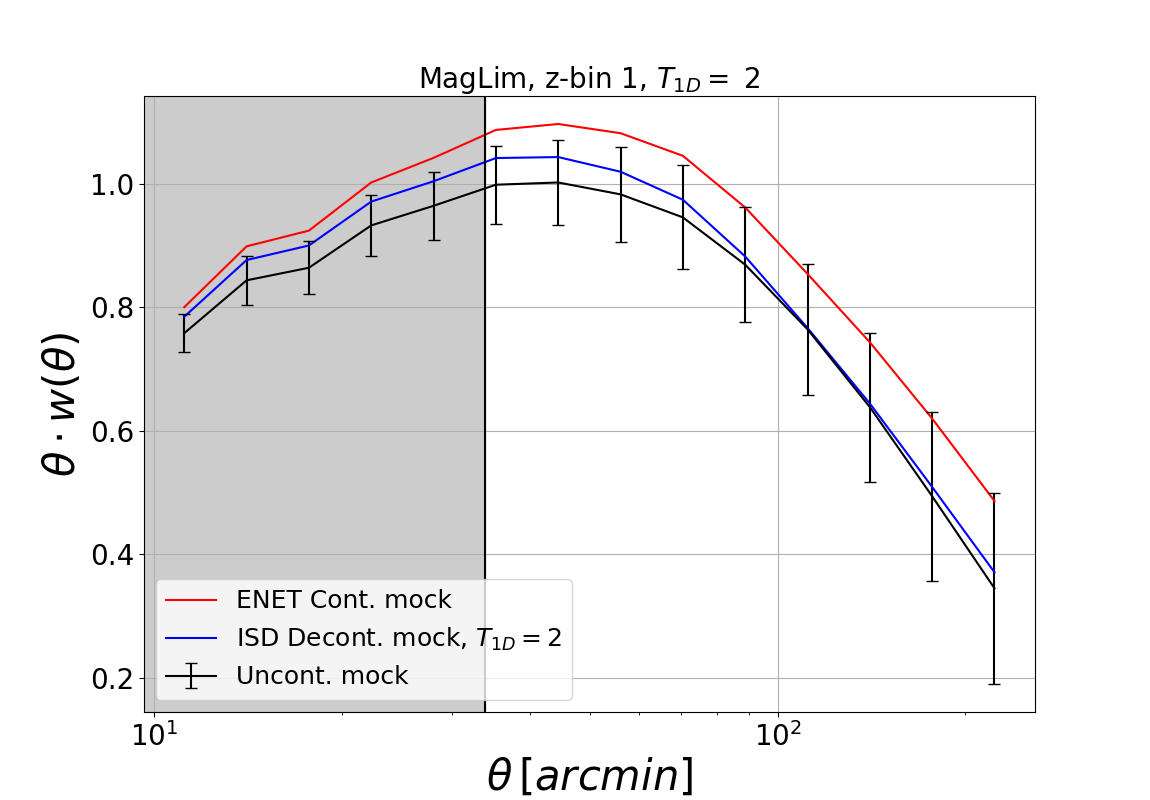}
    \includegraphics[width=\linewidth]{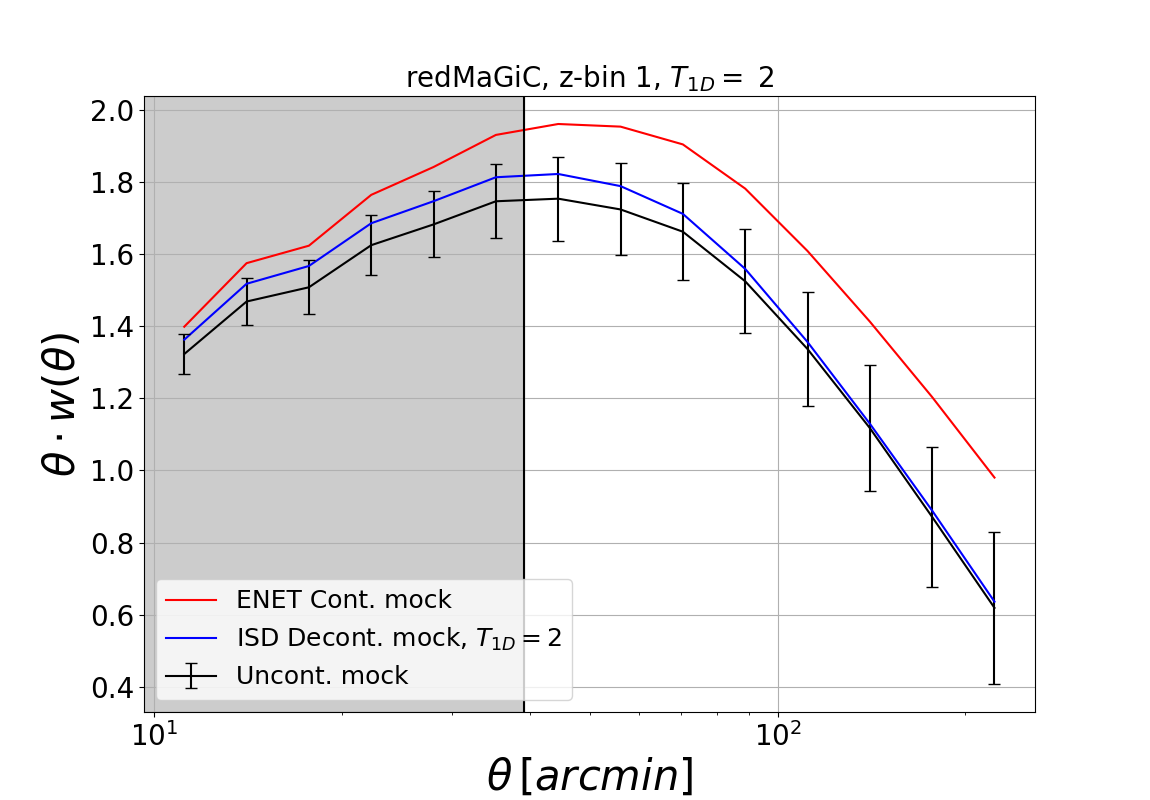}
    \caption{Mean angular correlation function, $w(\theta)$, from uncontaminated mocks (black line) and from decontaminated mocks (blue line) for \maglim (top panel) and for for \redmagic (bottom panel). The red line corresponds to the mean of the mocks with contamination added from \enet and the shaded regions represent the scales not used for cosmological constraints. While \isd recovers a nearly unbiased clustering at the largest angular scales, there is an important bias at the smallest ones. For this reason, this effect is marginalised over by adding it a systematic contribution to the error budget. The error bars shown take into account this contribution. }
    \label{fig:wtheta_resi_bias}
\end{figure}

In Figure \ref{fig:resi_bias}, we show the results for this bias with respect to the diagonal of the unaltered analytical errors. While the highest redshift bins of both \maglim and \redmagic present moderate levels of overcorrection, the lowest redshift bins of the two samples show a trend to under-correct at the small angular scales, but still above the scales we exclude. As already mentioned, we expect some level of undercorrection due to the aggressive contamination imprinted on the mocks. Even under this consideration, these bins cause the $\chi^2$ of the joint fit to exceed our limit, so we incorporate this bias as a systematic contribution to our covariance matrix. This is covered in Section \ref{sec:cov_sys_terms}. In Figure \ref{fig:wtheta_resi_bias}, we depict the mean recovered clustering (blue lines) compared to the true clustering (black lines). We also show the mean contaminated correlation function (red lines). It can be seen that \isd performs a nearly unbiased decontamination at the largest angular scales. The error bars in this Figure include the systematic terms added to the covariance (see Figure \ref{fig:method_diff_plot} for a comparison of the error bars with and without the systematic contributions).

\subsection{Impact on parameter estimation}\label{sec:impact_on_parameters}
\begin{figure*}
    \centering
    \includegraphics[width=0.49\linewidth]{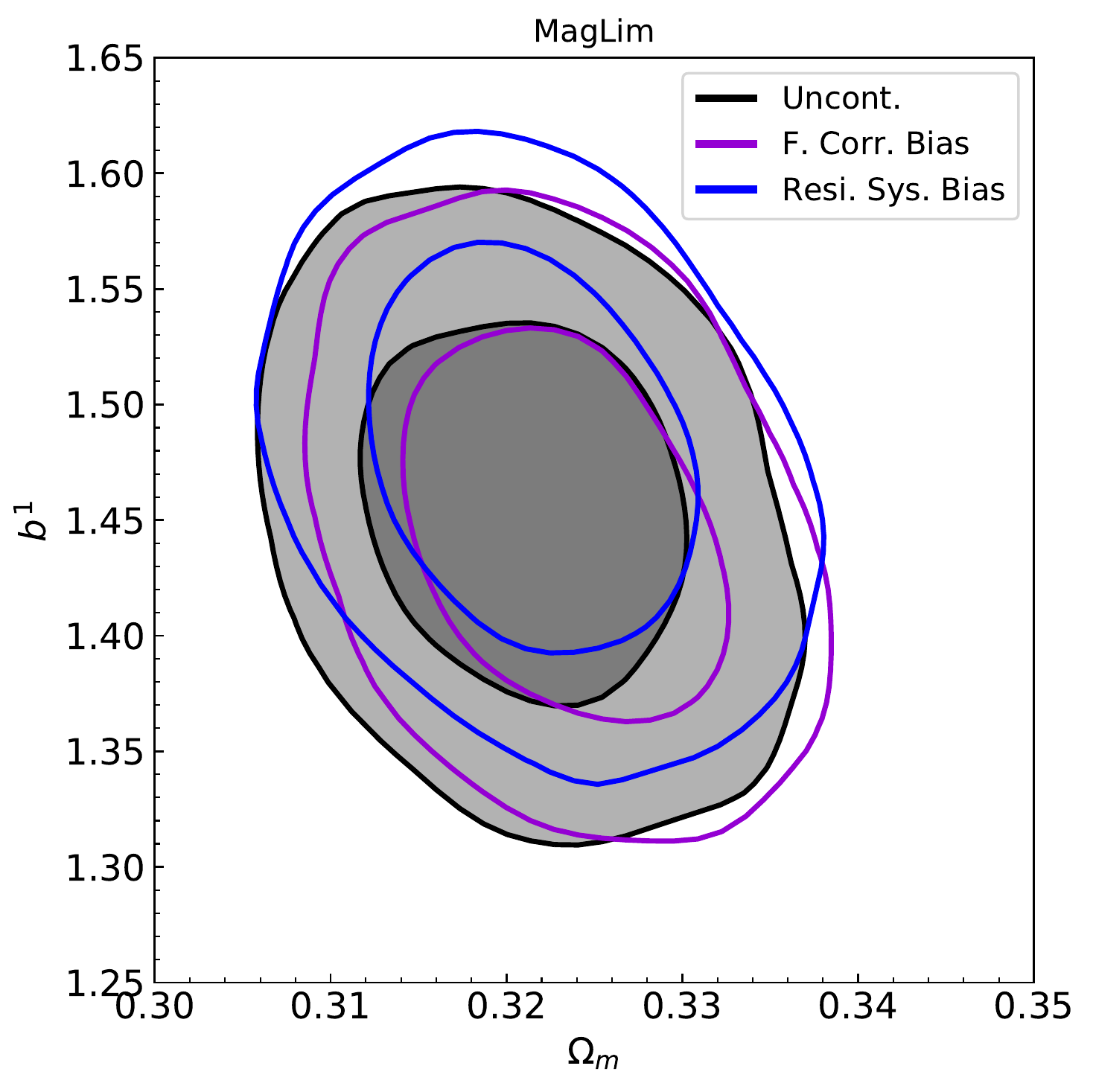}
    \includegraphics[width=0.49\linewidth]{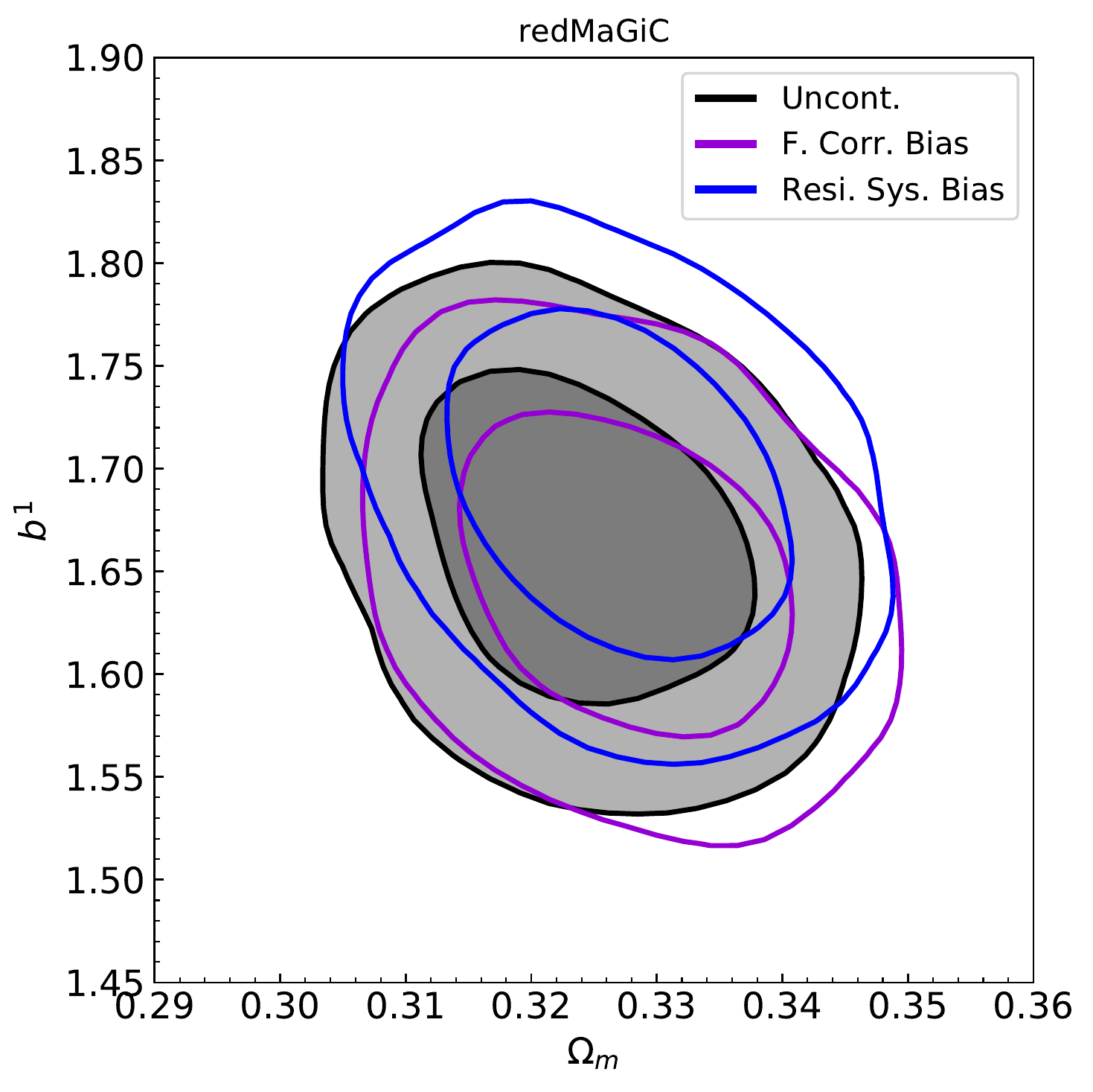}
    \caption{Constraints in the $\Omega_m \, - \, b^i$ parameter space at fixed $\sigma_8$ from the mean $w(\theta)$ of uncontaminated mocks (black contours) and from decontaminated mocks according to the false correction bias (violet contours) and to the residual systematic bias (blue contours). \maglim is shown in the left panel and \redmagic in the right one. It can be seen how both the false correction bias and the residual systematic bias lead to small shifts from the reference mocks relative to the error given by the \cosmolike \, analytical covariance, which includes the systematic uncertainty contributions. We only show contours for the first redshift bins of the two galaxy samples in this figure, but we verify that the shifts at the other bins are smaller or smaller. Because $\sigma_8$ and other cosmological parameters are fixed in this test, the posterior is smaller than from any of the DES final cosmological analyses that use the $w(\theta)$ data. }
    \label{fig:bias_parameters}
\end{figure*}

Finally, as an additional evidence of robustness we check the impact of the decontamination procedure on the estimation of cosmological parameters. We use as data vectors i) the mean correlation function over 400 uncontaminated mocks, ii) the mean correlation function biased by our overcorrection estimate (Section \ref{sec:false_correction_bias}) and iii) the mean correlation function biased as by the residual systematic uncertainty estimate (Section \ref{sec:residual_systematic_bias}). To test the influence of these analysis modifications on cosmology, we recalculate the constraints on the parameters $\Omega_m$ and $b^i$, marginalizing as before over redshift-bin centroid positions and widths of the redshift distributions. We use the same priors from Table \ref{tab:priors_info} and the rest of the parameters are fixed to the values used to generate the mocks. The results that we obtain are shown in Figure \ref{fig:bias_parameters}. It can be seen that the recovered contours from the false correction bias case (run on uncontaminated mocks) are in good agreement with those from the reference case, demonstrating that biases from overcorrection in inferred cosmological parameters are negligible. The contours corresponding to the residual systematic bias (run on \enet contaminated mocks) show a small level of undercorrection that is translated to slightly higher galaxy bias values, though this mismatch is also within the statistical uncertainties given by our analytical covariance. This covariance includes a systematic uncertainty correction that is explained in Section \ref{sec:cov_sys_terms}. In Table \ref{tab:mock_post_diff}, we present the difference in the $\Omega_m$ and $b^i$ mean posteriors in units of $\sigma$ from uncontaminated mock contours. We note that all differences are smaller than $0.5 \sigma$. It must be taken into account that, since the rest of the cosmological parameters are fixed, the $1\sigma$ contours are smaller than for any of the final DES cosmology analyses, making this test more stringent. We found that the mean $w(\theta)$ of the log-normal mocks is slightly shifted to lower amplitudes from the theory prediction with the same input values. This causes some shifting of the contours as well, but we have verified that this does not affect our conclusions from the decontamination methodology. 

\begin{table}
    \centering
    \begin{tabular}{c|c|c}
        \multicolumn{3}{ |c| }{\maglim } \\ \hline
        Parameter & False correction bias & Residual systematic bias \\ \hline 
        $\Omega_m$ & 0.36 & 0.08 \\ \hline 
        $b^1$ & -0.09 & 0.43 \\ \hline 
        $b^2$ & -0.06 & 0.40 \\ \hline 
        $b^3$ & -0.25 & 0.12 \\ \hline 
        $b^4$ & 0.05 & 0.16 \\ \hline 
        $b^5$ & -0.15 & -0.02 \\ \hline 
        $b^6$ & -0.06 & -0.04 \\ \hline 
        \hline
        \multicolumn{3}{ |c| }{\redmagic } \\ \hline
        Parameter & False correction bias & Residual systematic bias \\ \hline 
        $\Omega_m$ & 0.39 & 0.31 \\ \hline 
        $b^1$ & -0.29 & 0.50 \\ \hline 
        $b^2$ & -0.33 & 0.11 \\ \hline 
        $b^3$ & -0.30 & 0.27 \\ \hline 
        $b^4$ & -0.32 & -0.35 \\ \hline 
        $b^5$ & -0.19 & -0.21 \\ \hline 
        \hline
    \end{tabular}
    \caption{Relative difference in the $\Omega_m$ and $b^i$ mean of the posteriors for the two tests on decontaminated mocks in units of $\sigma$. All values are below half a $\sigma$. Note that the posteriors in this test are much smaller than in any of the final DES cosmology analyses because all the other parameters are fixed. }
    \label{tab:mock_post_diff}
\end{table}

\section{Post-unblinding investigations of the impact of observational systematics on \boldmath{$\texorpdfstring \MakeLowercase{w}(\theta)$}}\label{sec:redmagic_discussion}
The DES \threextwo analysis combines the correlation functions from galaxy clustering, $w(\theta)$, galaxy-galaxy lensing (for short, gg-lensing), $\gamma_t (\theta)$ and cosmic-shear, $\xi_{\pm} (\theta)$, in order to improve the individual constraining powers of each probe and to break degeneracies in some cosmological parameters. In addition, since each of these 2pt functions is potentially affected by different systematic effects, it allows for consistency checks comparing different results. The consideration of two different lens galaxy samples for $w(\theta)$ and $\gamma_t (\theta)$ allows us to further assess the robustness of the whole cosmology analysis. The cosmology analysis is performed blindly, that is, we only look at the cosmology results once a set of predefined criteria are fulfilled, as is described in \cite{y3-3x2ptkp}. During the unblinding process of \redmagic we found that this sample passed all the consistency tests we had a priori decided were required for unblinding. However, after unblinding, we identified a potential inconsistency between the amplitudes of galaxy clustering and gg-lensing: either the former has an anomalously high amplitude or the latter has an anomalously low one. This inconsistency is explored in detail in~\cite{y3-2x2ptbiasmodelling}.

Observational systematics from survey properties tend to increase the amplitude of $w(\theta)$ and so one possible explanation is that the clustering amplitude is anomalously high due to the decontamination procedure failing to fully capture all contamination in the data. Thus, the true underlying galaxy correlation function in the data would not be correctly recovered. This led us to perform a variety of additional tests as we describe below. It was during these tests when some of the methods described in Sections \ref{sec:sp_maps} and \ref{sec:methods} were incorporated, such as the change in SP map basis (both expanding the number of SP maps and decorrelating them) and the robustness checks using \enet and the neural net. Ultimately, we found that the difference between galaxy clustering and lensing observables in \redmagic remained robust to different choices in the decontamination procedure. We also applied these additional tests to the \maglim sample before it was unblinded. In contrast to our results with the \redmagic sample, once we unblinded the \maglim sample we found that its lensing and clustering signals were consistent with one another. For this reason, \maglim is the fiducial choice for our cosmological constraints \citep{y3-3x2ptkp}. The fiducial \maglim cosmology results use only the first four redshift bins, as the two highest redshift bins gave inconsistent results, while adding little constraining power. \cite{y3-2x2ptaltlensresults} investigates these results in detail.

\subsection{\isd and \enet at the STD map basis}\label{sec:stdmapbasis}
Before unblinding, \isd weights were obtained from a selection of STD maps performed by setting a limit for the Pearson's correlation coefficient between them. This selection gave 34 representative STD maps that were used to obtain weights with \isd (\isd -STD34). More details on this selection can be found in Appendix B of \cite{y3-baosample}. To check whether the clustering-lensing inconsistency found in \redmagic was caused by an STD map not selected in the STD34 set, we ran \isd on the full list of STD maps, and verified that derived weights did not significantly impact the resulting clustering signal. In Figure \ref{fig:wtheta_comparison}, we show the correlation functions at the first bin of \redmagic obtained for these two configurations of \isd with STD maps. 

We also checked the subtle possibility of a combination of STD maps leading to a large systematic contribution despite no single map being individually significant. For this reason, we ran \enet -STD107 on \redmagic , which simultaneously fits to all template maps, finding a significant decrease of $\sim 1\sigma$ in the amplitude of the correlation function in the first three redshift bins. This motivated further investigation to determine whether there could be significant residual contamination in the form of low-significance linear combinations of SP maps that eluded the initial decontamination procedure. We found that decorrelating the SP maps via PCA before running the \isd method and using the 107 components resulted in much better agreement between \isd and \enet, which motivated the change to the PC basis that has been used for the results presented in this paper (see \isd -PC107 in Figure \ref{fig:wtheta_comparison}). We also found that there are no significant changes when running \enet on the PC basis of maps (this method is less basis-dependent, since it performs a simultaneous fit to all maps).

\subsection{\isd and \enet in the PC map basis}
We evaluated the impact of the \isd-PC107 weights on both uncontaminated and \enet contaminated mocks, similar to the tests from Sections \ref{sec:false_correction_bias} and \ref{sec:residual_systematic_bias}. These tests revealed a significant level of overcorrection when using the full list of PC maps with \isd, especially when evaluated on contaminated mocks, indicating that true LSS fluctuations were being removed in the decontamination process. This effect can be seen in Figure \ref{fig:overcorrection_pc107}. We observed a similar overcorrection effect on \maglim with these \isd settings. The overcorrection is most prominent in lower redshift bins where the intrinsic clustering signal is larger, losing significance at higher redshift for both samples. 

These results suggest that there is a higher likelihood of chance correlation in the PC107 basis than in the STD107 basis. We also found that PC107 weights obtained from the data  showed significant correlations with DES $\kappa$ maps (see Appendix \ref{app:pc_cutoff} for details). We therefore conclude that using all 107 principal components results in removing not only actual systematic contamination from the data, but also cosmological signal, causing a lower $w(\theta)$ amplitude.

\begin{figure}
    \centering
    \includegraphics[width=\linewidth]{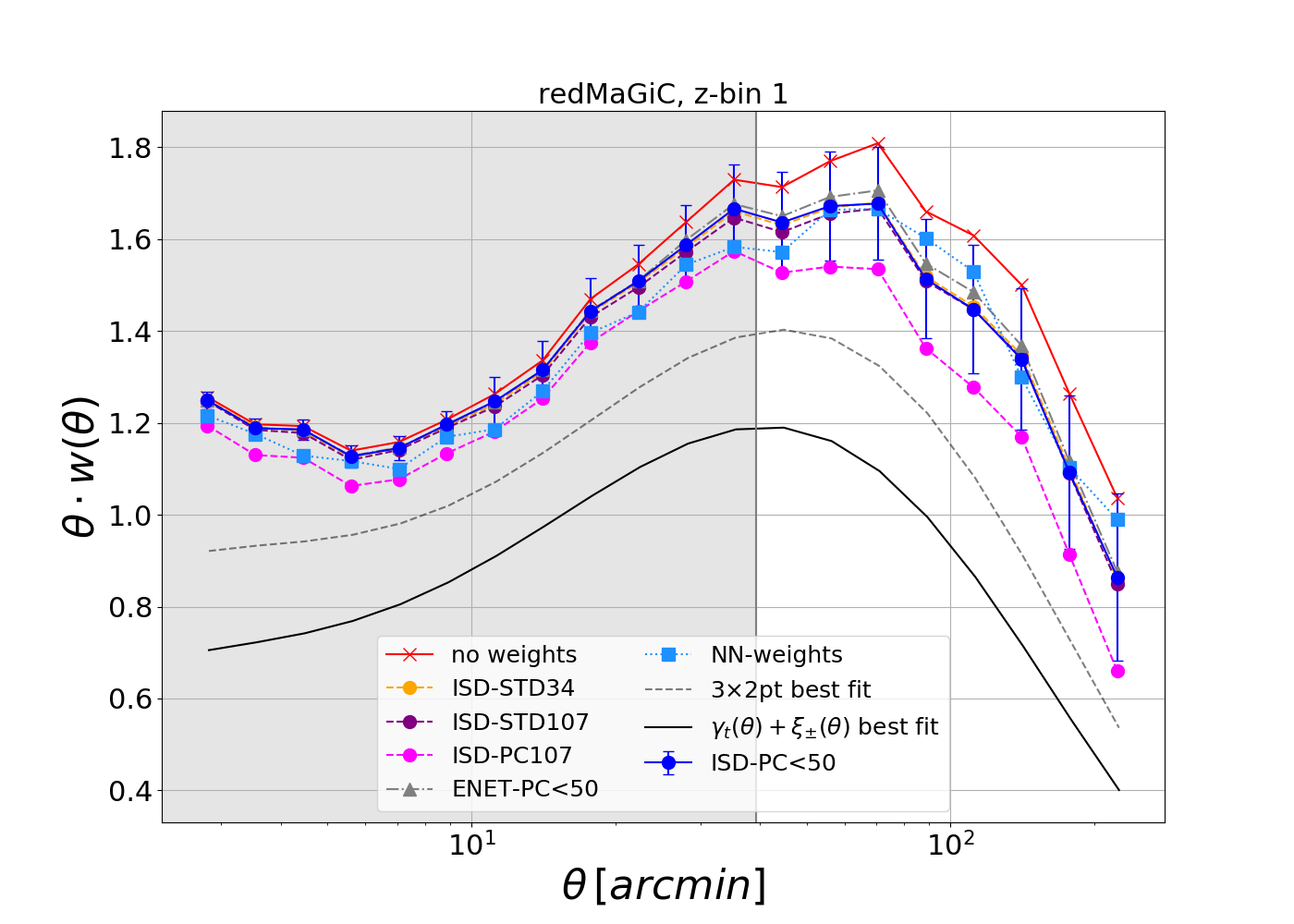}
    \caption{Comparison of the clustering amplitude recovered from several methods and configurations for the first redshift bin of \redmagic. All methods agree within the statistical uncertainty given by the analytical covariance. The solid red line corresponds to the unweighted data and the dashed purple line corresponds to the \isd -PC107 configuration. The difference between this configuration and the rest of methods is consistent with the overcorrection observed on contaminated mocks (see Figure \ref{fig:overcorrection_pc107}). The solid and dashed black lines are the best fit cosmology from cosmic-shear and gg-lensing only and from the 3$\times$2pt analysis, respectively. The gray shaded region represents the scales that are not used for cosmological analysis. None of the various configurations produce values of $w(\theta)$ approaching the best-fit prediction from cosmic-shear and gg-lensing.}
    \label{fig:wtheta_comparison}
\end{figure}

\begin{figure}
    \includegraphics[width=\linewidth]{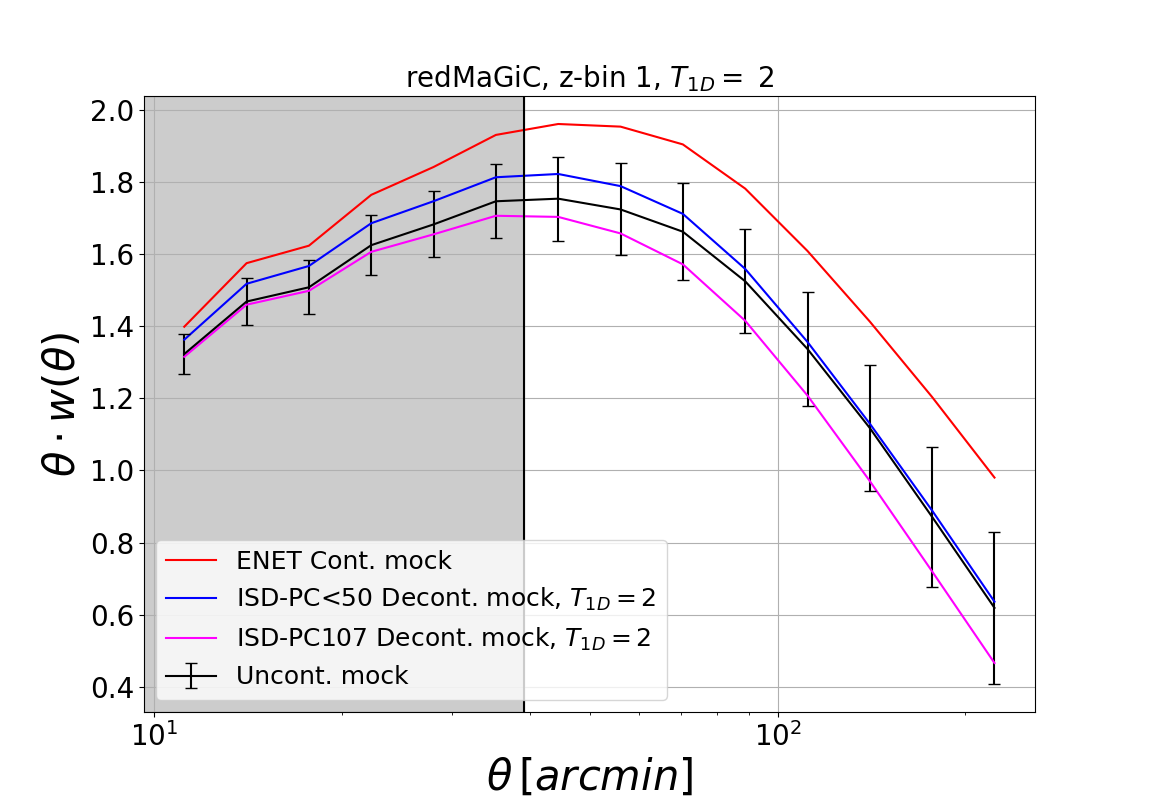}
    \caption{Effect of considering different numbers of PC maps on the two-point angular correlation function: weights obtained from 107 PC maps cause overcorrection on $w(\theta)$ (magenta line). This overcorrection ranges from $\sim 0.5 - 1\sigma$ and is most prominent at large angular scales. This overcorrection can explain most of the difference in clustering between \isd -PC<50 and \isd -PC107 observed in Fig.~\ref{fig:wtheta_comparison}. On the other hand, weights obtained from the first 50 PC maps yield a clustering amplitude (blue line) that is in good agreement with the mean $w(\theta)$ from uncontaminated mocks (black line), especially at the largest scales. The difference between the amplitudes from uncontaminated and \isd-PC<50 decontaminated mocks is included as a systematic contribution to the covariance (error bars in this figure already include that term). Red line correspond to the \enet-STD107 contaminated mocks.}
    \label{fig:overcorrection_pc107}
\end{figure}

We therefore applied a cutoff to the number of PC maps to be used. To select this cutoff, we required that the weight map resulting from running \isd with the set of the first $n$ PC maps should not induce a significant overcorrection on contaminated mocks (as we observed with \isd-PC107 weights), while still removing the contamination that was applied using \enet-STD107. We found that $n = 50$ principal component maps meets this requirement. The impact of the \isd-PC<50 weights on contaminated mocks and finally on the data can be seen in Figures \ref{fig:overcorrection_pc107} (blue line) and \ref{fig:wtheta_comparison} respectively. Then, we calculated \enet-PC<50 weights as well, finding good agreement between the two methods with this configuration (see Figure \ref{fig:wtheta_comparison}). Our adoption of this configuration was further supported by the desire to have a comparatively small number of maps to avoid overcorrection, as with the 107 PC maps, while still preserving most of the variance present in the full set of 107 STD maps. We point the reader to Appendix \ref{app:pc_cutoff} for more details on the selection of this cutoff. We found that the difference between $w(\theta)$ functions given by \isd-PC<50 and \enet-PC<50 yields a $\chi^2$ for the joint fit to all redshift bins bigger than 3. Thus, we consider this difference as an additional systematic uncertainty to be marginalised over, similar to the difference between uncontaminated and decontaminated mocks from Section \ref{sec:residual_systematic_bias}. 

For these reasons, we used \isd-PC<50 as the fiducial correction method, as described in the previous sections of this paper. In Figure \ref{fig:wtheta_comparison}, we summarise the clustering amplitudes obtained from each of the methods and configurations described in the first redshift bin of \redmagic. None of the methods produce a $w(\theta)$ consistent with the best fit prediction from cosmic-shear and gg-lensing (solid black line). For reference, the dashed grey line shows the best fit prediction from the combined \threextwo analysis. 

The tests conducted to determine this cutoff were focused on the first redshift bin of \redmagic, but we verified that the impact of this choice on the rest of the bins is similar, although milder, since the overcorrection observed at higher bins is less significant. We also ran these tests on \maglim, obtaining similar conclusions for the same cutoff.

\subsection{Tests with neural net weights}\label{sec:nn_weights}
As noted in Section~\ref{sec:nn_weights}, we developed an independent, nonlinear correction method using neural networks.  This was applied post-unblinding to test the robustness of the weights, in particular to the assumption of linearity between galaxy number density and the systematic maps. If there is excess clustering due to nonlinear functions of the STD maps, then we expect it to be captured by the NN-weights.
Because of the significant time required to run the method, we did not subject it to the full extent of validation tests on contaminated and uncontaminated mocks as we did for the \isd and \enet methods.
However as Fig.~\ref{fig:wtheta_comparison} shows, the changes to $w(\theta)$ are small when using the NN-weights, suggesting that residual nonlinear contamination from the existing set of STD maps is not driving a spuriously high estimate of $w(\theta)$.

\subsection{Modifications to the covariance matrix}\label{sec:cov_sys_terms}
\begin{figure*}
    \centering
    \includegraphics[width=\linewidth]{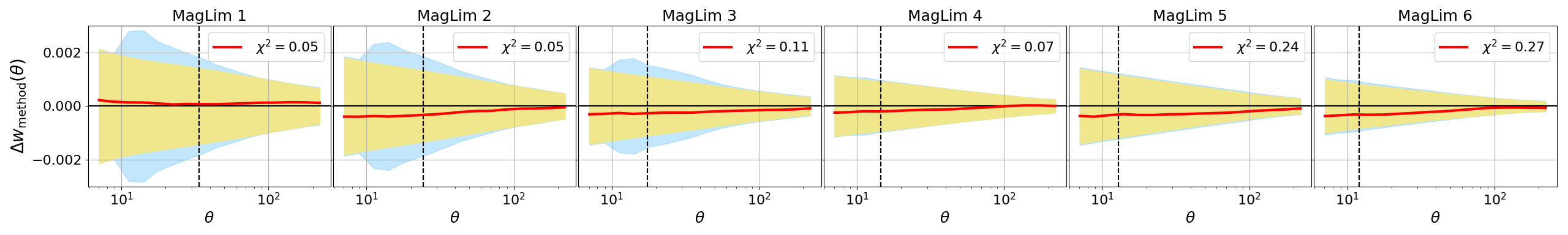}
    \includegraphics[width=0.83\linewidth]{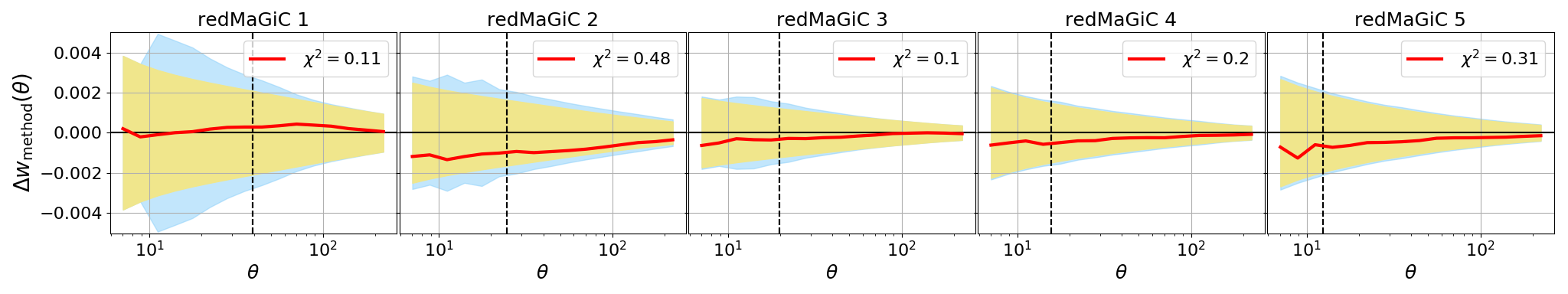}
    \caption{Method difference term $\Delta w_{\rm method}(\theta)$ in real data for \maglim (top row) and \redmagic (bottom row). The methods compared are \isd -PC<50 and \enet -PC<50 (red line). The light blue error bands correspond to the diagonal of the covariance with the additional systematic terms included, while the yellow ones correspond to the original analytical covariance.}
    \label{fig:method_diff_plot}
\end{figure*}

In this analysis, we consider the systematic uncertainty in the correction method from two sources: from the choice of correction method, and the bias measured in contaminated mocks (as mentioned in Section \ref{sec:residual_systematic_bias}). As noted in the previous section, the NN-weights method did not undergo the extensive validation process that the \isd and \enet weights did. For this reason, we focused on the systematic uncertainty associated to the differences between \isd-PC<50 and \enet-PC<50. \\ 
The two systematics considered are each analytically marginalised over through an additional term in the $w(\theta)$ covariance matrix following the methodology of \cite{bridle2002} summarised here. 
If one takes an arbitrary data vector ${\bf y}$ that is biased by an additive systematic effect ${\bf s}$,
\begin{equation}
{\bf y^\prime} = {\bf y} + A {\bf s},
\end{equation}
where A is the amplitude of the systematic error. If the amplitude $A$ has a Gaussian prior of zero-mean and width $\sigma_{A}$ (which can be determined by external constraints), the parameter A can be analytically marginalised over in the covariance matrix of ${\bf y}$ with, 
\begin{equation}
{\rm Cov}({\bf y^\prime}, {\bf y^\prime}) =  {\rm Cov}({\bf y}, {\bf y}) + \sigma_{A}^2 {\bf s}{\bf s}^{T} .
\end{equation} 
In this analysis, we model the impact of the systematic uncertainty in the correction as, 
\begin{equation}
w^\prime(\theta) = w(\theta) + A_{1} \Delta w_{\rm method}(\theta) + A_{2} w^{T_{1D}}_{\rm r. \, s. \, bias}(\theta) ,
\end{equation}
where $\Delta w_{\rm method}(\theta)$ is the difference between the \isd and \enet methods, both using the PC<50 basis of maps as shown in Fig. \ref{fig:method_diff_plot}; $w^{T_{1D}}_{\rm r. \, s. \, bias}(\theta)$ is the residual systematic bias measured on Log-normal mocks in Sec. \ref{sec:residual_systematic_bias}, and $A_{1}$ and $A_{2}$ are two arbitrary amplitudes that describe the size of the systematic error in the correction. \\
\\
We analytically marginalise over these terms assuming a unit Gaussian as the prior on the amplitudes $A_{1}$ and $A_{2}$ such that the measured systematic size is a 1$\sigma$ deviation from the prior centre, and the systematic can move $w(\theta)$ in either direction. The final additional covariance term is 
\begin{equation}
\Delta {\rm Cov}( {\bf w^\prime}, {\bf w^\prime} ) =  {\bf \Delta w_{\rm method}} {\bf \Delta w_{\rm method}}^{T} \ + \ {\bf w^{T_{1D}}_{r. \, s. \, bias}} {\bf w^{T_{1D}}_{r. \, s. \, bias}}^{T} .
\end{equation} 
The method difference term $\Delta w_{\rm method}(\theta)$ is measured on real data and therefore contains the same noise as the $w(\theta)$ data vector being used for cosmological inference. To avoid adding this noise to the covariance term, we fit a flexible polynomial to the two $w(\theta)$ measurements described in Appendix \ref{app:polynomial_fit}. $\Delta w_{\rm method}(\theta)$ is the difference between these two polynomial fits.

The mock bias term $w^{T_{1D}}_{\rm r. \, s. \, bias}(\theta)$ is averaged over 400 mocks so is a smooth function of $\theta$ and does not require any additional fitting. The impact of the additional covariance terms is shown in the error bars of Fig. \ref{fig:method_diff_plot}. 
The systematic contribution to each tomographic bin is treated as independent so the covariance between bins is not modified. 

\subsection{Tests with \balrog }\label{sec:balrog_tests}
\begin{figure*}
    \centering
    \includegraphics[width=\linewidth]{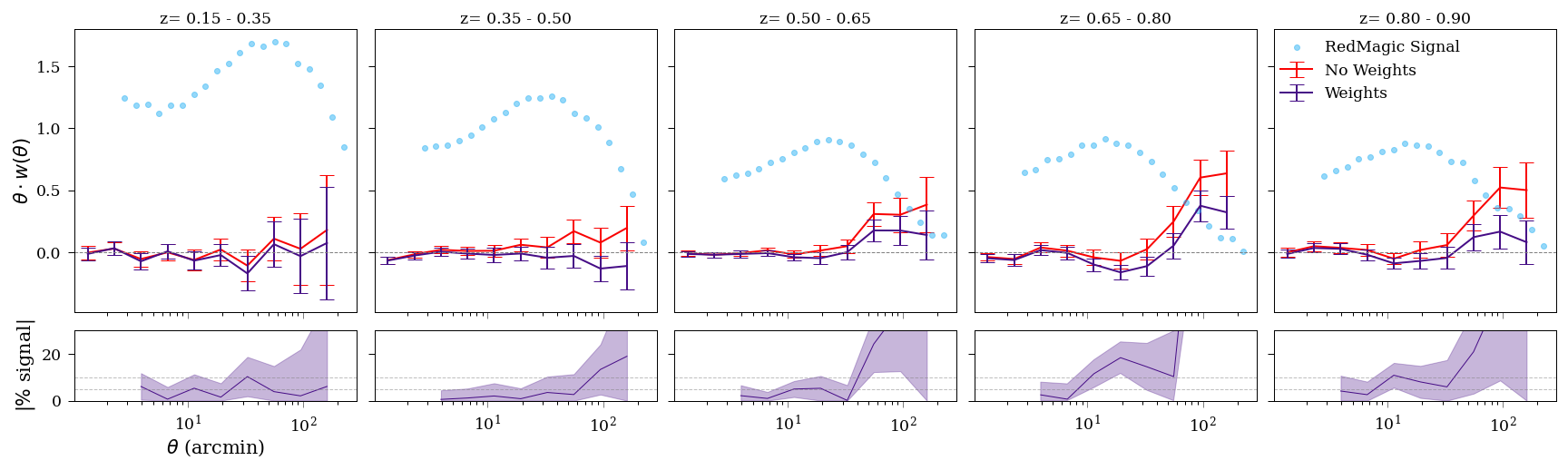}
    \includegraphics[width=\linewidth]{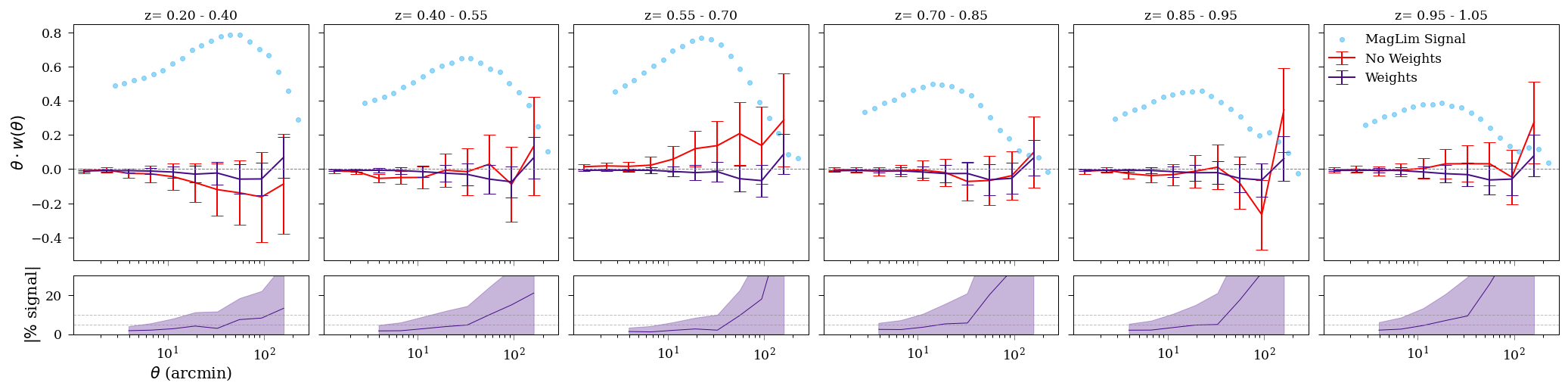}
    \caption{Cross-correlation between \redmagic (top panel) and \maglim (bottom panel) samples selected in data and produced with \balrog . The cross-correlations are shown in the top row of each panel, before weighting (red line) and after weighting (purple line) by SP maps effects, compared to the data $w(\theta)$ (blue points). The error bars have been obtained by jackknife re-sampling. The bottom row of each panel shows the relative difference (in percent) between the cross-correlation signal and the auto-correlation one. In general, all differences are compatible with zero and well below the statistical errors showing no clear evidence of uncorrected effects from imaging systematics, though we note that the points for \redmagic are noisier due to its lower number density.}
    \label{fig:cross_balrog}
\end{figure*}

\balrog \, \citep{2016MNRAS.457..786S, y3-balrog} is a software package which embeds fake objects in real images in order to accurately characterize measurement effects. It is a useful tool to make independent consistency tests of the decontamination methods. While the galaxy samples trace the actual large-scale structure, the \balrog \, samples are formed by galaxies that are artificially injected on a uniform grid. What both real and \balrog \, samples have in common is the impact of systematics. Therefore, any correlation between the two after applying the weights would mean the presence of a common systematic. For this reason, we used the cross-correlation of \redmagic and \maglim with their associated \balrog \, samples to test for the presence of an extraneous signal that would indicate a systematic which is not being corrected by the applied weights. These results are presented in Figure \ref{fig:cross_balrog}. The cross-correlations are calculated in $\sim 1000 \deg^2$ (available area of the \balrog \, samples). All errors (computed with jackknife re-sampling using 100 patches for \maglim and 50 for \redmagic) for the cross-correlation with the weights applied are consistent with zero signal. However, the signal itself is small but nonzero, growing in magnitude towards larger scales. We note that, due to its lower number density, the points for \redmagic are noisier than those for \maglim. The reduced $\chi^2$ for a constant cross-correlation of 0 are 0.46, 0.96, 1.25, 3.60, 1.18 for \redmagic and 1.13, 0.71, 0.78, 0.94, 0.65, 0.69 for \maglim. 
The relative strength of the cross-correlation signal with respect to the auto-correlation signal can be seen in the bottom rows of each panel. 
In general, it is at or below 5\% for the five lowest angular bins at all redshift bins, and it is lower than 10\% for scales smaller than $\sim 30$ arcmin. This relative strength gives us an indication of the size of a systematic effect that could be still unaccounted for. Even if the \redmagic results are noisy, those for \maglim do not show a clear indication of uncorrected effects from imaging systematics. 

\subsection{Summary of findings}\label{sec:wtheta_conclusions}
We performed a series of tests post-unblinding to determine if the observed inconsistency between the galaxy clustering and gg-lensing signals in \redmagic is due to residual systematic contamination of the galaxy clustering signal. In particular, we investigated whether expanding the set of survey property maps, adjusting the contamination model, or changing a variety of methodological choices for the decontamination procedure resulted in a significantly different inferred galaxy clustering signal. We largely performed these tests at the level of $w(\theta)$, without further looking at the impact of these decisions on cosmological parameters. The following list is a summary of the obtained results:

\begin{itemize}
    \item Expanding the list of 34 to all 107 STD maps has negligible impact on the resulting amplitude of $w(\theta)$ using the fiducial \isd decontamination procedure. We thus conclude that the discrepancy is not due to residual contamination from one of the previously-discarded STD maps.
    \item  We performed a principle component analysis of the 107 STD maps and used the principle components as an orthonormal basis for the decontamination procedure, i.e. ran \isd -PC107. We found good agreement with \enet -STD107 (and \enet -PC107), resulting in a reduction of the $w(\theta)$ amplitude. This was most pronounced in the first redshift bin of \redmagic, with a decrease in $w(\theta)$ of $\sim 1 \sigma$.
    \item We observed a significant overcorrection of $w(\theta)$ when computing \isd-PC107 weights from contaminated mocks. For this reason, we applied a cutoff to the number of PC maps, limiting it to the 50 PC maps with the highest signal-to-noise. We found that the resultant \isd-PC<50 weights produce little overcorrection and we add a systematic contribution to our error budget corresponding to the difference between \isd-PC<50 and \enet-PC<50. We also add a systematic contribution for the undercorrection observed on contaminated mocks using only the first 50 PC maps assuming the true contamination corresponds to the estimate of \enet -STD107.
    \item We implemented a nonlinear decontamination procedure using a neural network, which also used different choices for the mask and base set of STD maps. This resulted in differences in $w(\theta)$ that were much smaller than the observed discrepancy between galaxy clustering and gg-lensing. 
    \item We cross-correlated both \redmagic and \maglim with their corresponding \balrog \, samples and we found no clear evidence of uncorrected contamination of known systematic templates common to both types of samples. 
\end{itemize}

We note that the \isd -STD34 weights passed an extensive battery of validation tests, described in Section \ref{sec:weights_validation}. However, after our findings and comparisons between \enet and \isd we decided to use the \isd -PC<50 weights in the fiducial analysis.

Given these findings, we conclude that the anomalous high clustering amplitude of \redmagic sample is unlikely to be due to uncorrected contamination coming from any of our known templates nor from a linear combination of them. Because the clustering remains high when using higher-order STD maps with \enet (after accounting for false correction bias) as well as using the neural net, we are unable to identify non-linear contamination from our SP maps as the cause (see Appendix \ref{app:chi2_null_dist} for additional tests). We performed a number of further exploratory tests such as more aggressive masking, including based on the leverage statistic \citep[c.f.  ][]{Weaverdyck_2021} and found $w(\theta)$ to be robust to these choices. Applying our fiducial decontamination procedure to \maglim does not show the same discrepancy between probes as does \redmagic.

\section{Conclusions}\label{sec:conclusions}
We measure the angular two-point correlation of DES Y3 lens galaxies, and study the impact of systematic errors on these measurements. We use two lens samples: \maglim , a magnitude-limited sample with enhanced number density and reliable photometric redshifts \citep{y3-2x2maglimforecast} and \redmagic , a sample of luminous red galaxies (LRGs) selected by the algorithm described in \citet*{Rozo:2016} which also provides high quality photometric redshifts. We extend the methodology employed in DES Y1 \citep{Elvin-Poole:2017xsf}, both for correcting the data and to ensure its robustness. A more thorough set of survey property maps is used and we employ them directly and through the application of principal components analysis to the map set. Additionally, a new weight estimation method is used in parallel \citep[\enet,][]{Weaverdyck_2021} and a cross-check of linearity assumptions is made with a neural network framework based on recent literature \citep{bossnn}. These steps help us to avoid possible blind spots in our validation methodology.   

Our findings are as follows:
\begin{itemize}
\item The updated DES Y1 methodology, dubbed \textit{Iterative Systematics Decontamination} (\isd), is able to successfully remove systematic contamination, as shown by  validation tests on log-normal mocks (Figures \ref{fig:wtheta_false_corr_bias} and \ref{fig:wtheta_resi_bias}) and data. 
\item The \enet method is a viable alternative correction method to \isd . We evaluate several configurations and demonstrate that both methods are in agreement within statistical precision. To be sure that any residual difference is taken into account, we include a systematic uncertainty in the covariance matrix as the difference between the two results. This uncertainty is included in the final covariance that is used for cosmological constraints, after checking that it does not bias our results.
\item The decontamination procedure does not produce a significant bias in $w(\theta)$ or in the $\Omega_m - b^i$ parameter space. 
\item We find that survey properties have a significant impact on the recovered galaxy clustering signal, particularly at high redshifts, as compared to \redmagic Y1 results \citep{Elvin-Poole:2017xsf}. This contamination is corrected by applying the \isd method together with a principal component analysis of our survey sroperty maps. The same methodology is applied to both samples. 
\item We find an inconsistent clustering amplitude for the \redmagic sample when combined with other 2pt lensing probes. We study it from the point of view of the impact of SP maps, considering different methods, such as \isd and \enet , and different numbers, types and bases of SP maps. We find agreement between the weighted correlation functions yielded by each method within our errors. We also investigate weights from a neural network weighting scheme. All our tests confirm that our systematics corrections are robust and the template maps used in this analysis do not explain the \redmagic internal inconsistency. 
\end{itemize}

The results presented in this work have been optimized to be used for their combination with galaxy-galaxy lensing (\citealt{y3-2x2ptaltlensresults, y3-gglensing, y3-2x2ptbiasmodelling}; \citealt*{y3-2x2ptmagnification}) and cosmic-shear (\citealt{y3-cosmicshear1}; \citealt*{y3-cosmicshear2}) measurements to obtain the 3$\times$2pt cosmological results from the DES Year 3 data \citep{y3-3x2ptkp}, and constitutes one of the basic pillars for this measurement. 

This work highlights the importance of adequate validation and cross-checking of this highly relevant step in the estimation of galaxy clustering, and builds upon several developments within the DES project and in the literature. For Y6, given the rapid developments in the field, we plan to approach the problem from the beginning with a variety of methodologies in mind, possibly considering multi-regression approaches or assessing the feasibility of using a wider \balrog\, sample, making it part of the pipeline from the start now that the algorithm is fully developed. This will be coupled with possibly a multi-tiered unblinding approach with additional steps to be able to make decisions on investigating unusual results in internal consistency tests at different stages of the process. Additional work in parallel on the Y3 samples and survey property maps will shed some light on possible details that the Y6 methodology will have to address, such as understanding the overcorrection produced by some maps or issues with the galaxy samples.

\section{Acknowledgements}

Funding for the DES Projects has been provided by the U.S. Department of Energy, the U.S. National Science Foundation, the Ministry of Science and Education of Spain, 
the Science and Technology Facilities Council of the United Kingdom, the Higher Education Funding Council for England, the National Center for Supercomputing 
Applications at the University of Illinois at Urbana-Champaign, the Kavli Institute of Cosmological Physics at the University of Chicago, 
the Center for Cosmology and Astro-Particle Physics at the Ohio State University,
the Mitchell Institute for Fundamental Physics and Astronomy at Texas A\&M University, Financiadora de Estudos e Projetos, 
Funda{\c c}{\~a}o Carlos Chagas Filho de Amparo {\`a} Pesquisa do Estado do Rio de Janeiro, Conselho Nacional de Desenvolvimento Cient{\'i}fico e Tecnol{\'o}gico and 
the Minist{\'e}rio da Ci{\^e}ncia, Tecnologia e Inova{\c c}{\~a}o, the Deutsche Forschungsgemeinschaft and the Collaborating Institutions in the Dark Energy Survey. 

The Collaborating Institutions are Argonne National Laboratory, the University of California at Santa Cruz, the University of Cambridge, Centro de Investigaciones Energ{\'e}ticas, Medioambientales y Tecnol{\'o}gicas-Madrid, the University of Chicago, University College London, the DES-Brazil Consortium, the University of Edinburgh, the Eidgen{\"o}ssische Technische Hochschule (ETH) Z{\"u}rich, Fermi National Accelerator Laboratory, the University of Illinois at Urbana-Champaign, the Institut de Ci{\`e}ncies de l'Espai (IEEC/CSIC), the Institut de F{\'i}sica d'Altes Energies, Lawrence Berkeley National Laboratory, the Ludwig-Maximilians Universit{\"a}t M{\"u}nchen and the associated Excellence Cluster Universe, the University of Michigan, NFS's NOIRLab, the University of Nottingham, The Ohio State University, the University of Pennsylvania, the University of Portsmouth, SLAC National Accelerator Laboratory, Stanford University, the University of Sussex, Texas A\&M University, and the OzDES Membership Consortium.

Based in part on observations at Cerro Tololo Inter-American Observatory at NSF's NOIRLab (NOIRLab Prop. ID 2012B-0001; PI: J. Frieman), which is managed by the Association of Universities for Research in Astronomy (AURA) under a cooperative agreement with the National Science Foundation.

The DES data management system is supported by the  National Science Foundation under Grant Numbers AST-1138766 and AST-1536171.
The DES participants from Spanish institutions are partially supported by MICINN under grants ESP2017-89838, PGC2018-094773, PGC2018-102021, SEV-2016-0588, SEV-2016-0597, and MDM-2015-0509, some of which include ERDF funds from the European Union. IFAE is partially funded by the CERCA program of the Generalitat de Catalunya.
Research leading to these results has received funding from the European Research Council under the European Union's Seventh Framework Program (FP7/2007-2013) including ERC grant agreements 240672, 291329, and 306478.
We  acknowledge support from the Brazilian Instituto Nacional de Ci\^encia e Tecnologia (INCT) do e-Universo (CNPq grant 465376/2014-2).

This manuscript has been authored by Fermi Research Alliance, LLC under Contract No. DE-AC02-07CH11359 with the U.S. Department of Energy, Office of Science, Office of High Energy Physics.

\bibliographystyle{mnras_2author}
\bibliography{systematics,des_y3kp} 

\section*{Affiliations}
$^{1}$ Centro de Investigaciones Energ\'eticas, Medioambientales y Tecnol\'ogicas (CIEMAT), Madrid, Spain\\
$^{2}$ Department of Physics, University of Michigan, Ann Arbor, MI 48109, USA\\
$^{3}$ Center for Cosmology and Astro-Particle Physics, The Ohio State University, Columbus, OH 43210, USA\\
$^{4}$ Department of Physics, The Ohio State University, Columbus, OH 43210, USA\\
$^{5}$ Institut d'Estudis Espacials de Catalunya (IEEC), 08034 Barcelona, Spain\\
$^{6}$ Institute of Space Sciences (ICE, CSIC),  Campus UAB, Carrer de Can Magrans, s/n,  08193 Barcelona, Spain\\
$^{7}$ Instituto de Astrofisica de Canarias, E-38205 La Laguna, Tenerife, Spain\\
$^{8}$ Laborat\'orio Interinstitucional de e-Astronomia - LIneA, Rua Gal. Jos\'e Cristino 77, Rio de Janeiro, RJ - 20921-400, Brazil\\
$^{9}$ Universidad de La Laguna, Dpto. Astrofísica, E-38206 La Laguna, Tenerife, Spain\\
$^{10}$ Instituto de F\'{i}sica Te\'orica, Universidade Estadual Paulista, S\~ao Paulo, Brazil\\
$^{11}$ Instituto de Fisica Teorica UAM/CSIC, Universidad Autonoma de Madrid, 28049 Madrid, Spain\\
$^{12}$ Physics Department, 2320 Chamberlin Hall, University of Wisconsin-Madison, 1150 University Avenue Madison, WI  53706-1390\\
$^{13}$ Department of Physics and Astronomy, University of Pennsylvania, Philadelphia, PA 19104, USA\\
$^{14}$ Department of Physics, Northeastern University, Boston, MA 02115, USA\\
$^{15}$ Laboratory of Astrophysics, \'Ecole Polytechnique F\'ed\'erale de Lausanne (EPFL), Observatoire de Sauverny, 1290 Versoix, Switzerland\\
$^{16}$ Lawrence Berkeley National Laboratory, 1 Cyclotron Road, Berkeley, CA 94720, USA\\
$^{17}$ Department of Physics, Carnegie Mellon University, Pittsburgh, Pennsylvania 15312, USA\\
$^{18}$ NSF AI Planning Institute for Physics of the Future, Carnegie Mellon University, Pittsburgh, PA 15213, USA\\
$^{19}$ Santa Cruz Institute for Particle Physics, Santa Cruz, CA 95064, USA\\
$^{20}$ Department of Astronomy/Steward Observatory, University of Arizona, 933 North Cherry Avenue, Tucson, AZ 85721-0065, USA\\
$^{21}$ Institute of Theoretical Astrophysics, University of Oslo. P.O. Box 1029 Blindern, NO-0315 Oslo, Norway\\
$^{22}$ Jet Propulsion Laboratory, California Institute of Technology, 4800 Oak Grove Dr., Pasadena, CA 91109, USA\\
$^{23}$ Kavli Institute for Cosmology, University of Cambridge, Madingley Road, Cambridge CB3 0HA, UK\\
$^{24}$ Institut de F\'{\i}sica d'Altes Energies (IFAE), The Barcelona Institute of Science and Technology, Campus UAB, 08193 Bellaterra (Barcelona) Spain\\
$^{25}$ Center for Astrophysical Surveys, National Center for Supercomputing Applications, 1205 West Clark St., Urbana, IL 61801, USA\\
$^{26}$ Department of Astronomy, University of Illinois at Urbana-Champaign, 1002 W. Green Street, Urbana, IL 61801, USA\\
$^{27}$ Department of Astronomy, University of Geneva, ch. d'\'Ecogia 16, CH-1290 Versoix, Switzerland\\
$^{28}$ Fermi National Accelerator Laboratory, P. O. Box 500, Batavia, IL 60510, USA\\
$^{29}$ Department of Applied Mathematics and Theoretical Physics, University of Cambridge, Cambridge CB3 0WA, UK\\
$^{30}$ Kavli Institute for Particle Astrophysics \& Cosmology, P. O. Box 2450, Stanford University, Stanford, CA 94305, USA\\
$^{31}$ Kavli Institute for the Physics and Mathematics of the Universe (WPI), UTIAS, The University of Tokyo, Kashiwa, Chiba 277-8583, Japan\\
$^{32}$ Department of Astronomy and Astrophysics, University of Chicago, Chicago, IL 60637, USA\\
$^{33}$ Kavli Institute for Cosmological Physics, University of Chicago, Chicago, IL 60637, USA\\
$^{34}$ ICTP South American Institute for Fundamental Research\\ Instituto de F\'{\i}sica Te\'orica, Universidade Estadual Paulista, S\~ao Paulo, Brazil\\
$^{35}$ Department of Physics, University of Arizona, Tucson, AZ 85721, USA\\
$^{36}$ SLAC National Accelerator Laboratory, Menlo Park, CA 94025, USA\\
$^{37}$ Department of Physics, Stanford University, 382 Via Pueblo Mall, Stanford, CA 94305, USA\\
$^{38}$ Institute of Cosmology and Gravitation, University of Portsmouth, Portsmouth, PO1 3FX, UK\\
$^{39}$ Institute for Astronomy, University of Hawai'i, 2680 Woodlawn Drive, Honolulu, HI 96822, USA\\
$^{40}$ CNRS, UMR 7095, Institut d'Astrophysique de Paris, F-75014, Paris, France\\
$^{41}$ Sorbonne Universit\'es, UPMC Univ Paris 06, UMR 7095, Institut d'Astrophysique de Paris, F-75014, Paris, France\\
$^{42}$ Department of Physics and Astronomy, Pevensey Building, University of Sussex, Brighton, BN1 9QH, UK\\
$^{43}$ Department of Physics \& Astronomy, University College London, Gower Street, London, WC1E 6BT, UK\\
$^{44}$ Jodrell Bank Center for Astrophysics, School of Physics and Astronomy, University of Manchester, Oxford Road, Manchester, M13 9PL, UK\\
$^{45}$ University of Nottingham, School of Physics and Astronomy, Nottingham NG7 2RD, UK\\
$^{46}$ Astronomy Unit, Department of Physics, University of Trieste, via Tiepolo 11, I-34131 Trieste, Italy\\
$^{47}$ INAF-Osservatorio Astronomico di Trieste, via G. B. Tiepolo 11, I-34143 Trieste, Italy\\
$^{48}$ Institute for Fundamental Physics of the Universe, Via Beirut 2, 34014 Trieste, Italy\\
$^{49}$ Observat\'orio Nacional, Rua Gal. Jos\'e Cristino 77, Rio de Janeiro, RJ - 20921-400, Brazil\\
$^{50}$ Department of Physics, IIT Hyderabad, Kandi, Telangana 502285, India\\
$^{51}$ Institute of Astronomy, University of Cambridge, Madingley Road, Cambridge CB3 0HA, UK\\
$^{52}$ School of Mathematics and Physics, University of Queensland,  Brisbane, QLD 4072, Australia\\
$^{53}$ Center for Astrophysics $\vert$ Harvard \& Smithsonian, 60 Garden Street, Cambridge, MA 02138, USA\\
$^{54}$ Australian Astronomical Optics, Macquarie University, North Ryde, NSW 2113, Australia\\
$^{55}$ Lowell Observatory, 1400 Mars Hill Rd, Flagstaff, AZ 86001, USA\\
$^{56}$ Departamento de F\'isica Matem\'atica, Instituto de F\'isica, Universidade de S\~ao Paulo, CP 66318, S\~ao Paulo, SP, 05314-970, Brazil\\
$^{57}$ George P. and Cynthia Woods Mitchell Institute for Fundamental Physics and Astronomy, and Department of Physics and Astronomy, Texas A\&M University, College Station, TX 77843,  USA\\
$^{58}$ Department of Astrophysical Sciences, Princeton University, Peyton Hall, Princeton, NJ 08544, USA\\
$^{59}$ Department of Astronomy, University of Michigan, Ann Arbor, MI 48109, USA\\
$^{60}$ Instituci\'o Catalana de Recerca i Estudis Avan\c{c}ats, E-08010 Barcelona, Spain\\
$^{61}$ Faculty of Physics, Ludwig-Maximilians-Universit\"at, Scheinerstr. 1, 81679 Munich, Germany\\
$^{62}$ Max Planck Institute for Extraterrestrial Physics, Giessenbachstrasse, 85748 Garching, Germany\\
$^{63}$ School of Physics and Astronomy, University of Southampton,  Southampton, SO17 1BJ, UK\\
$^{64}$ Computer Science and Mathematics Division, Oak Ridge National Laboratory, Oak Ridge, TN 37831\\
$^{65}$ Universit\"ats-Sternwarte, Fakult\"at f\"ur Physik, Ludwig-Maximilians Universit\"at M\"unchen, Scheinerstr. 1, 81679 M\"unchen, Germany\\

\appendix
\section{Log-normal mocks}\label{app:app_mocks}
The mocks used for the systematic analysis are 2D log-normal fields generated at a given power spectrum. We start by using \Camb \, \citep{Lewis:1999bs,Howlett:2012mh} to obtain a matter power spectrum and project into a galaxy clustering angular power spectrum, $C^{gg}_i(l)$ following the theory modelling described in \cite{y3-generalmethods}. To produce this power spectrum we assume our fiducial cosmology and we fix the galaxy bias for each redshift bin to the values from the blind bias analysis (Table \ref{tab:samples_info}). Then, we use this power spectrum to generate a Gaussian random field of $\delta_g$ for each mock realization on a \healpix \, map \citep{healpix} using the \healpy package \citep{Zonca2019}. We then apply a log-normal transformation to the field following the methodology of \cite{Xavier:2016}. This uses a skewness parameter which was derived in \cite{y3-covariances}. We then transform the log-normal $\delta_g$ field to a galaxy number counts field, $N_{\rm gal}$,  
using the observed number count, $\Bar{N}_{\rm o}$, from the galaxy sample we want to reproduce and the relation, 
\begin{equation}
    N_{\rm gal} = \Bar{N}_{\rm o} \cdot (1+\delta_g) \, .
\end{equation} 
We apply the angular mask to the full sky realizations. In this way, the covariance matrices built from these mocks incorporate the same mask effects as the real data. In order to add shot noise, we finally Poisson sample the $N_{\rm gal}$ field. 

\begin{figure}
    \centering
    \includegraphics[width=\linewidth]{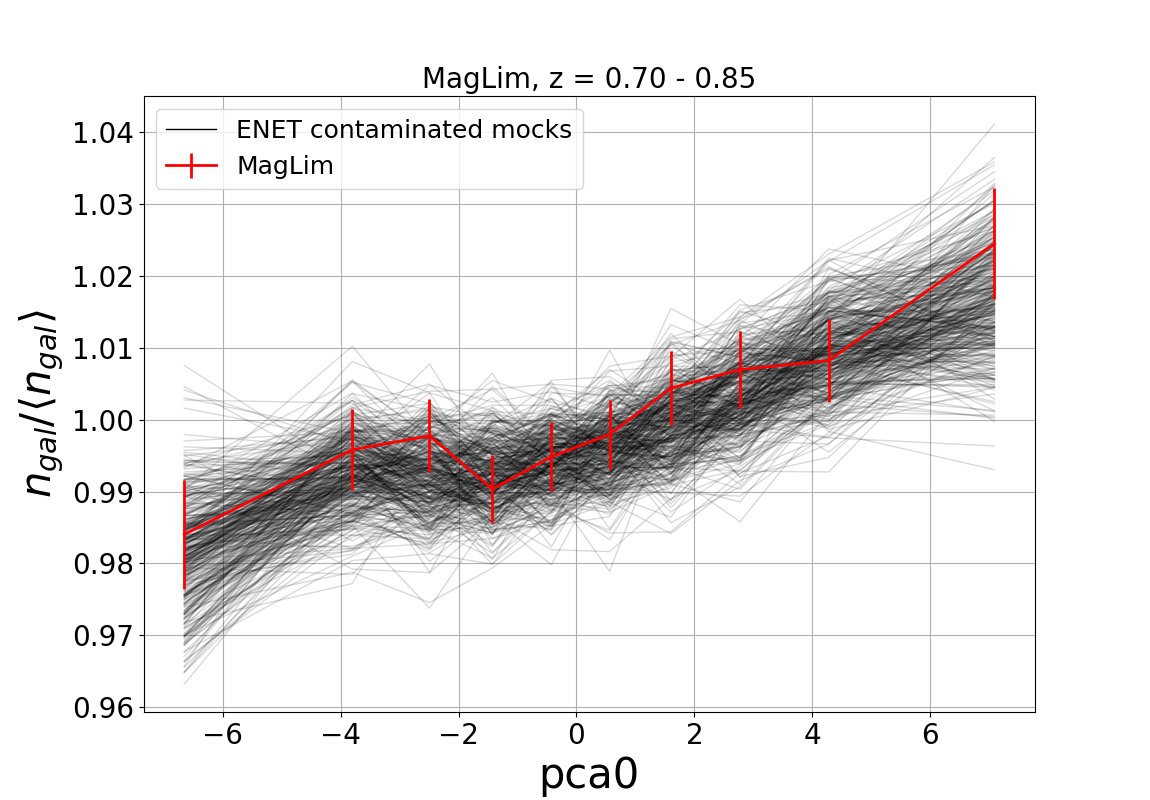}
    \includegraphics[width=\linewidth]{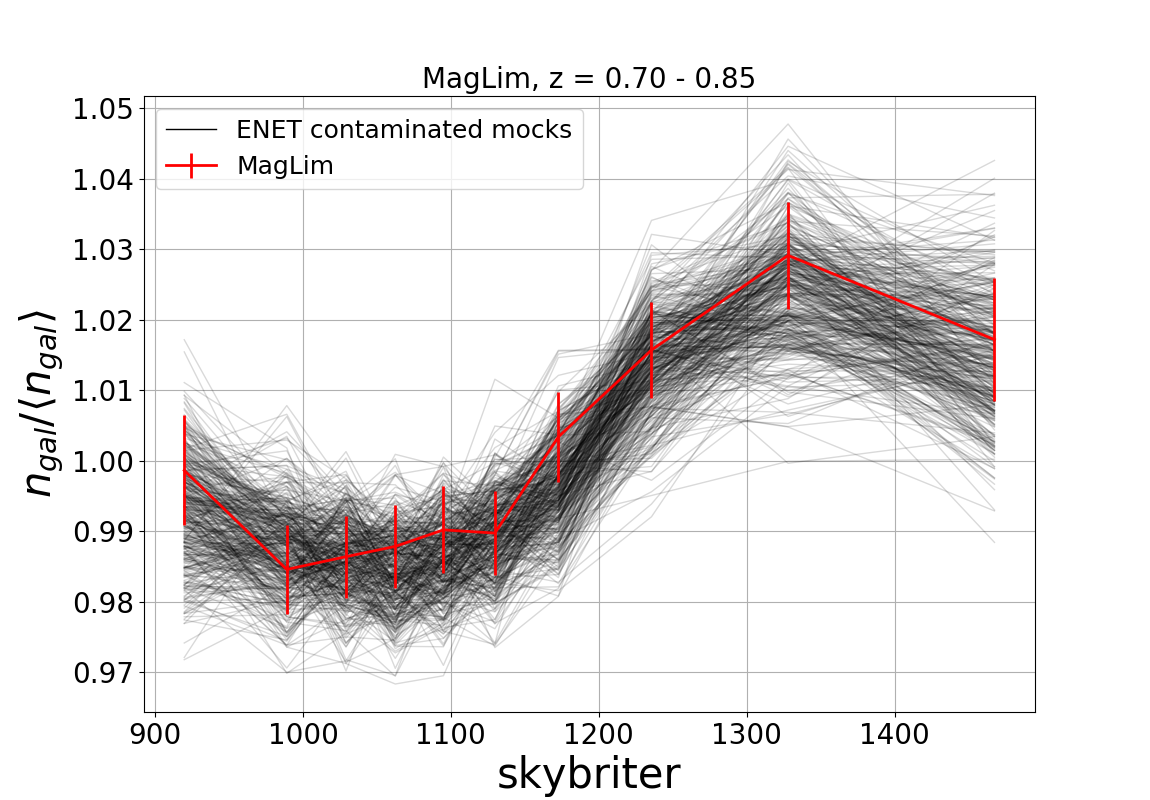}
    \caption{1D relations for 400 \maglim \enet contaminated log-normal mocks (shaded black lines) compared with the data (red line). The top panel shows the 1D relations with the \pcmap{0} map at the fourth redshift bin of this sample, whereas the bottom panel shows the 1D relations with \skybrite \, in $r$-band. The contamination observed on the data is well reproduce by these mocks. The error bars are obtained from the uncontaminated mocks used to calculate the 1D significance.}
    \label{fig:test1}
\end{figure}

As we mention in Section \ref{sec:weights_validation}, we also create a set of log-normal mocks contaminated by survey properties systematics, so we can look for biases introduced by \isd and check their impact on the measurements. 
We imprint contamination on the log-normal mocks by multiplying the galaxy number counts field by the inverse of the weight map derived from the data, that is
\begin{equation}
    N_{\rm gal, \, mock}^p \rightarrow N_{\rm gal, \, mock}^p \cdot \frac{1}{w^p} \, .
\end{equation}\\
This step is applied before Poisson sampling the galaxy field. We produce a set of 400 contaminated log-normal mocks following this procedure using weights derived from \enet -STD107, as is mentioned in Section \ref{sec:weights_validation}. We check that the 1D relations of these mocks reproduce in shape and amplitude those observed on the data. An example of this can be seen in Figure \ref{fig:test1}.

\section{Internal consistency tests: Estimator bias test}\label{app:estimator_bias}

\begin{figure}
    \centering
    \includegraphics[width=\linewidth]{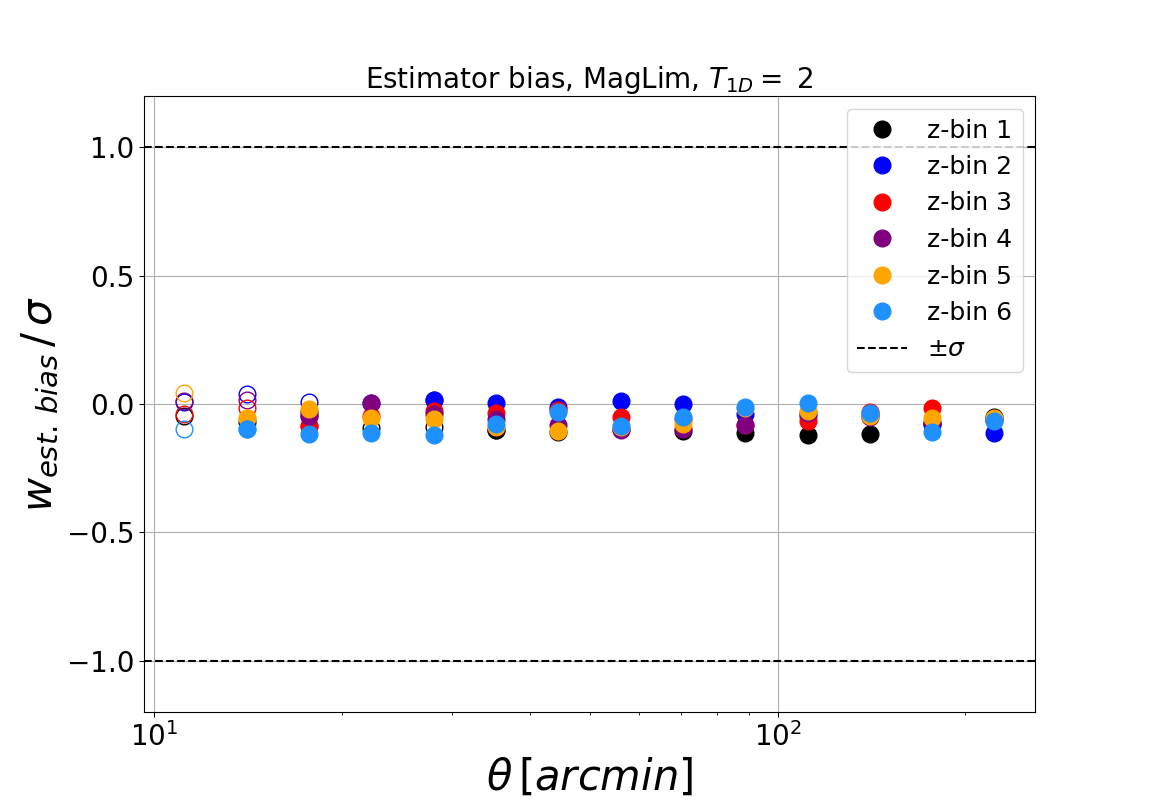}
    \includegraphics[width=\linewidth]{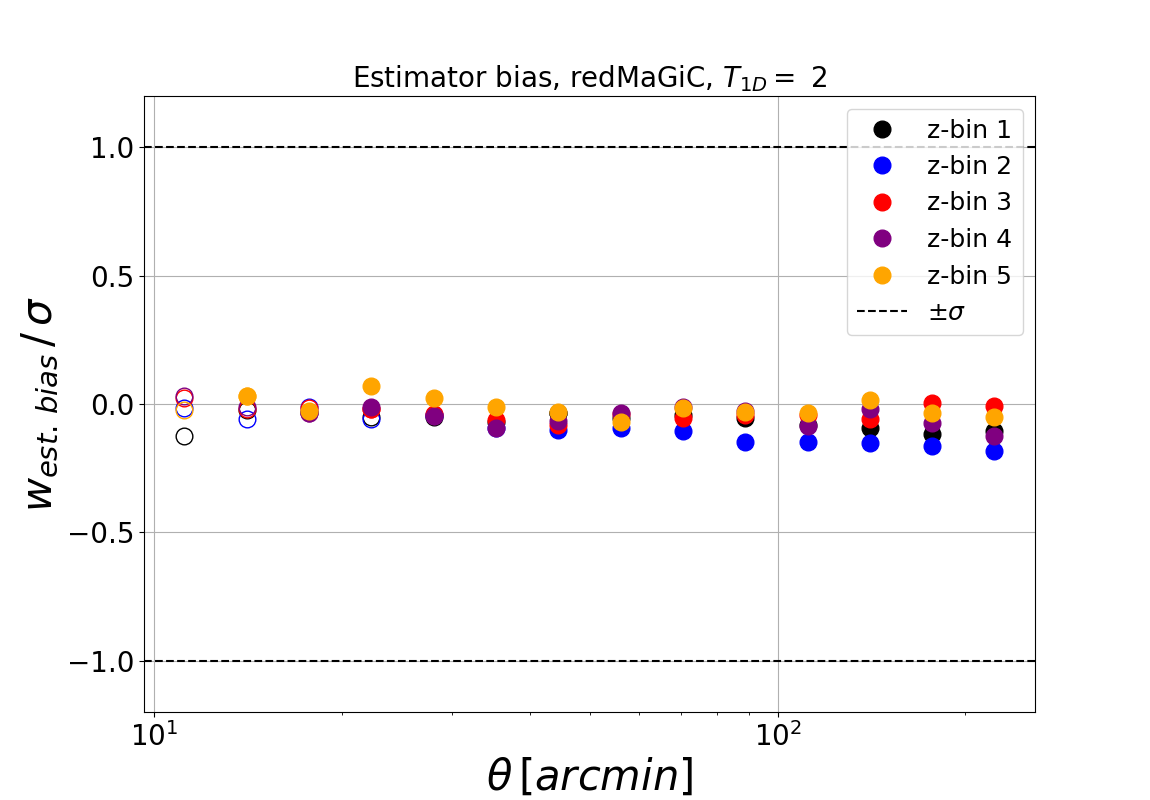}
    \caption{Estimator bias for \maglim(top panel) and \redmagic(bottom panel). The negative values are due to small level of overcorrection. Empty dots correspond to the scales excluded for each redshift bin. }
    \label{fig:est_bias}
\end{figure}

In addition to the tests described in Section \ref{sec:weights_validation}, we perform an internal consistency test that seeks to confirm no bias in $w(\theta)$ is introduced by \isd under idealised circumstances. For this test we contaminate and correct for the same list of SP maps, demonstrating the Landy-Szalay estimator can recover a negligibly biased signal.  
Since the focus of this test is the $w(\theta)$ estimator itself when applied to weighted data, independently of the origin of these weights, we conduct it using weights from a preliminary run of \isd on the standard SP maps, with the same threshold that we use to obtain the weights from the data, $T_{1D} = 2$. To get the magnitude of this potential bias, we defined
\begin{equation}
w_{\rm est. \, bias}(\theta) = \frac{1}{N} \, \left( \sum_{i=n} ^N w_{\rm dec, \, \textit{i}} \, - \, \sum_{j=1} ^N  w_{\rm unc, \, \textit{j}} \right)(\theta) 
\end{equation}

\noindent where $w_{\rm unc, \, \textit{i}}$ are the correlation functions from uncontaminated mocks, $w_{\rm dec, \, \textit{i}}$ are those from decontaminated mocks and $N = 1000$ mock realizations. Figure \ref{fig:est_bias} showcases the values of $w_{\rm est. \, bias}(\theta)$. As it can be seen, we see no indication of estimator bias for both lens samples at every redshift bin. This demonstrates that the combination of our weighting methodology with the Landy-Szalay estimator for $w(\theta)$ does not induce any bias on our measurements when the list of contaminating SP maps is known. 

\section{Polynomial fits for \boldmath{$\Delta \texorpdfstring \MakeLowercase{w_{\rm method}}(\theta)$}}\label{app:polynomial_fit}
\begin{figure}
    \centering
    \includegraphics[width=\linewidth]{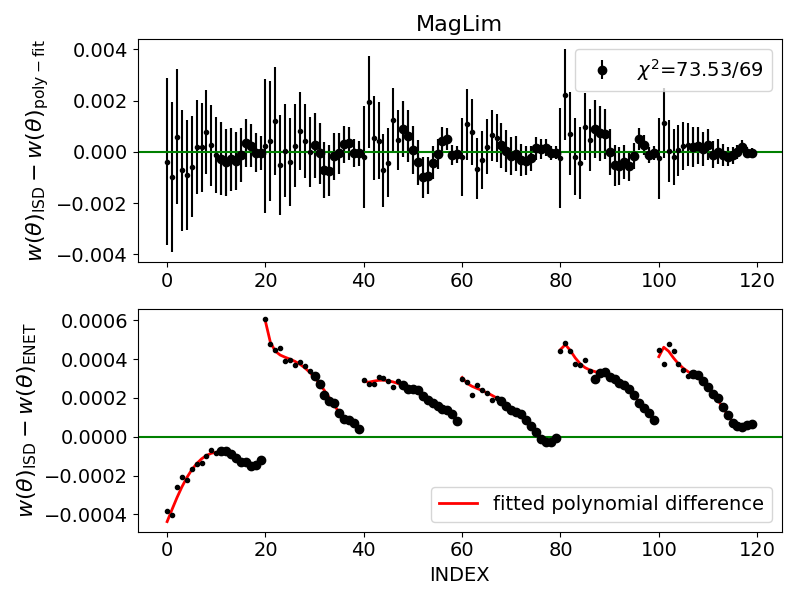}
    \includegraphics[width=\linewidth]{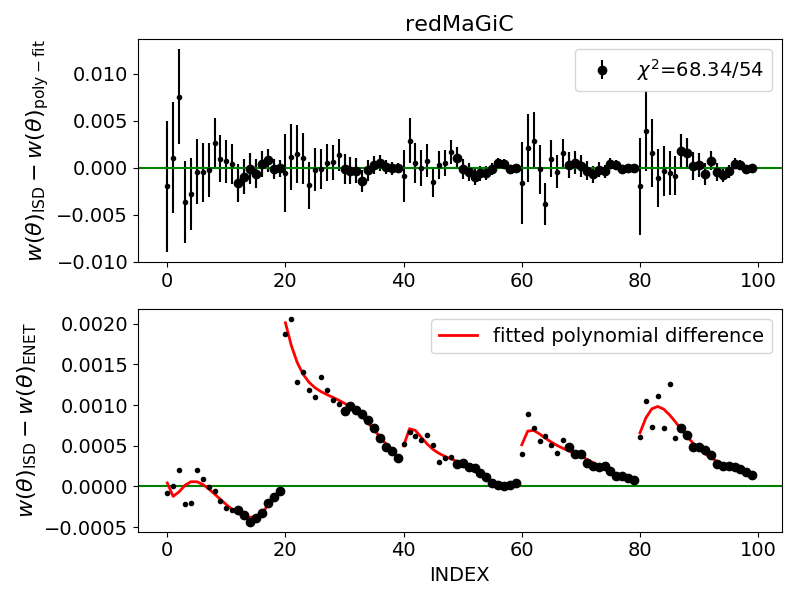}
    \caption{ Polynomial fits to $w(\theta)$ data used in estimating the systematic terms in the $w(\theta)$ covariance in Sec.~\ref{sec:cov_sys_terms}. The first and third panels show the fit residuals to the fiducial $w(\theta)$ measurements for each sample. The second and fourth panels show the difference between the polynomial fits of the two correction methods considered in these terms, \isd and \enet both with the first 50 principle component template maps. The bold points are the data included by the scale cuts and included in the fit and $\chi^2$ calculations. }
    \label{fig:poly_fit_best_fit}
\end{figure}

The additional covariance term described in Sec.~\ref{sec:cov_sys_terms} depends on the difference between $w(\theta)$ measured with two different systematics correction methods, $\Delta w_{\rm method}(\theta)$. As $\Delta w_{\rm method}(\theta)$ is measured on real data, it  contains the same noise as the $w(\theta)$ data vector being used for cosmological inference. To avoid adding this noise to the covariance term, we fit a flexible polynomial to the two $w(\theta)$ measurements in the form, 
\begin{equation}
w_{\rm poly fit}(\theta) = \sum_{i=-3}^{+3} B_{i} \theta^{i}
\label{eq:poly_fit}
\end{equation}
where $B_{i}$ are the coefficients to be fitted. 
The best fit polynomials are shown in Fig.~\ref{fig:poly_fit_best_fit}. We find this polynomial to be a good fit to the data, and the difference between measured correlation functions matches the difference in fitted polynomials well. 

\section{Principal Component maps cutoff}\label{app:pc_cutoff}

\begin{figure}
    \centering
    \includegraphics[width=\linewidth]{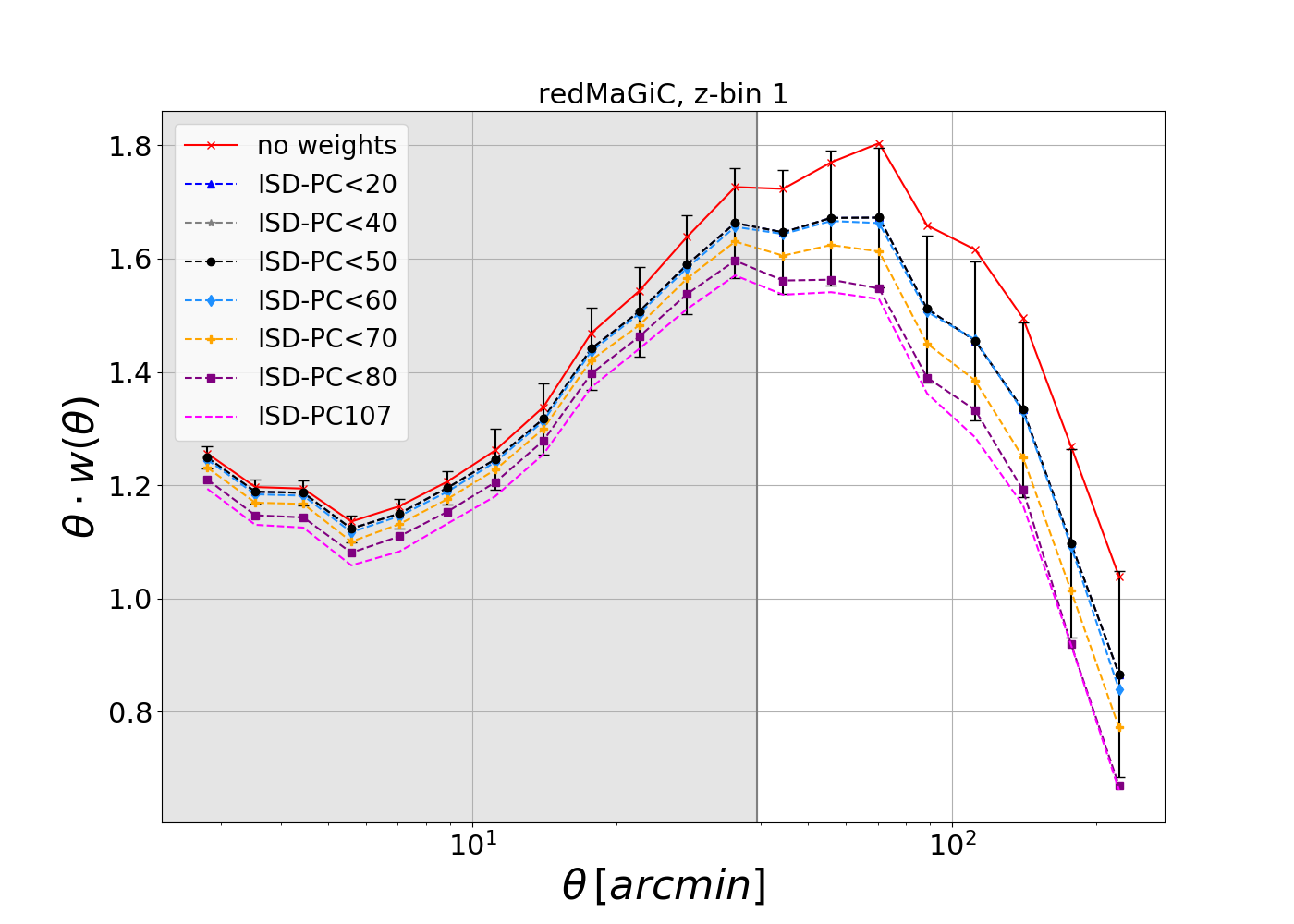}
    \caption{Clustering amplitude at the first redshift bin of \redmagic for several PC cutoffs, \isd -PC<$n$. The solid red line corresponds to the unweighted data and the dashed magenta one to the weights obtained from \isd -PC107, which lead to overcorrection. It can be seen how around $n=50$ the $w(\theta)$ amplitudes converge. }
    \label{fig:wtheta_pc_cutoffs}
\end{figure}

In Sec.~\ref{sec:redmagic_discussion} we describe a set of systematics weights using only the first 50 principle component maps labelled \isd -PC<50, which are used as the fiducial weights in the cosmology analysis. In this appendix we provide some further justification for this choice.

\begin{figure*}
    \centering
    \includegraphics[width=\linewidth]{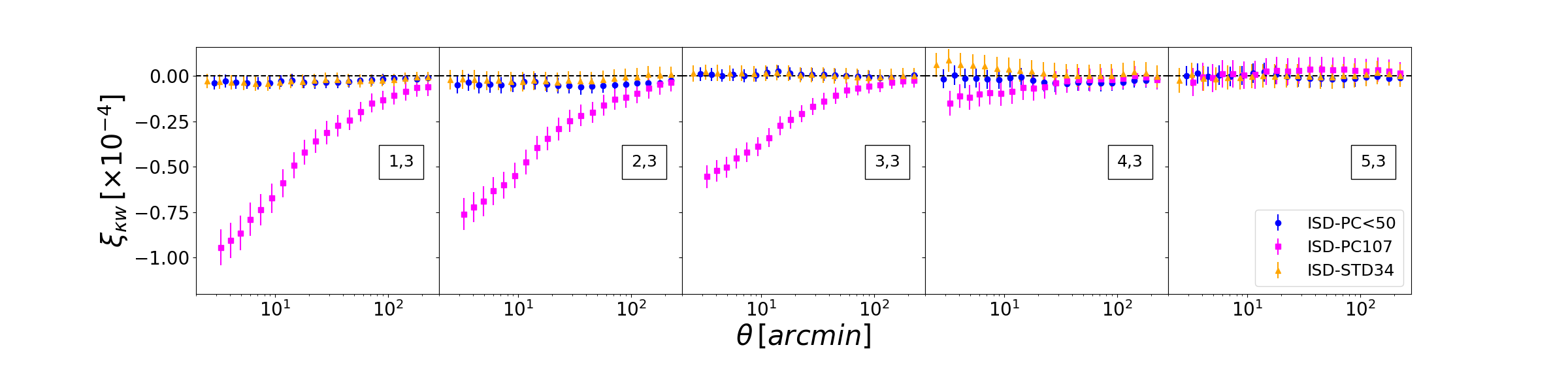}
    \caption{Cross-correlation of weight maps from different configurations of \isd with the convergence field, $\kappa$. The error-bars are calculated using jackknife with 150 patches. It can be seen how the \isd -PC107 weights cross-correlate significantly with $\kappa$, while the weights from the other two configurations do not. This suggests that the high PC template maps may correlate with LSS. An off-set has been added to the x-axis points for better visualisation. }
    \label{fig:kappa_cross}
\end{figure*}

In order to test for the correlation of real large-scale structure with the weight maps, we cross-correlate the convergence, $\kappa$, maps from \citet*{y3-massmapping} with the weight maps obtained using different methods, \isd -STD34, \isd -PC107 and \isd -PC<50. We correlate with the convergence map for the third tomographic source bin due to the large overlap between its lensing kernel and the lens sample. In the absence of systematics in the $\kappa$ maps, we do not expect there to be correlations between the SP or weight maps and the convergence maps. We show these correlations in Fig.~\ref{fig:kappa_cross} for the five \redmagic tomographic bins (the error bars are estimated using jackknife methodology using 150 patches). We find that while \isd run on only the 34 representative STD maps does not correlate with the convergence maps, we obtain a large correlation with the weight maps using all the PC maps, pointing to potential leakage of cosmological structure in these weights, either from chance correlation or real large-scale structure leaking into the high PC maps. 

\begin{figure*}
    \centering
    \includegraphics[width=\linewidth]{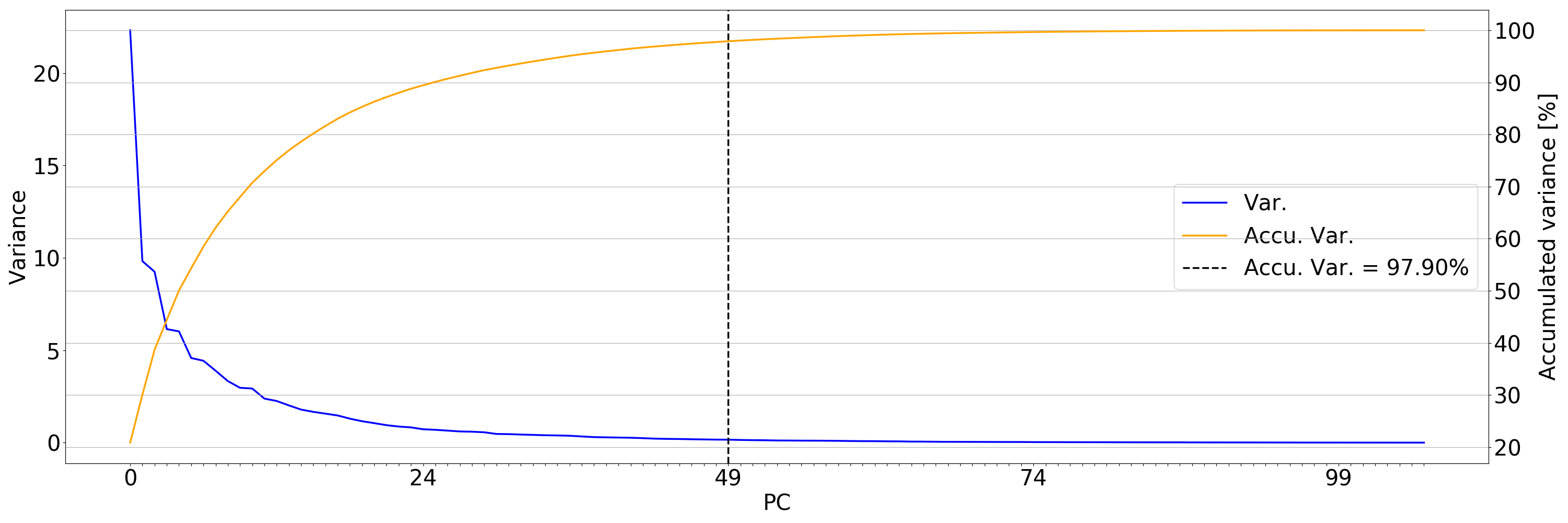}
    \caption{Variance of each PC map (blue line) and percent of accumulated variance (orange line). For the principal component map 49 the accumulated variance is $\sim 98\%$, so the remaining maps are compatible with noise. }
    \label{fig:scree_plot}
\end{figure*}

To mitigate any correlation with real large-scale structure, we restrict the weight estimation to use only the first $n$ PC maps. 
First of all, to ensure that all dominant features of the SP maps are taken into account, we look at the amount of variance captured up to each component. This is shown on Figure \ref{fig:scree_plot}. Based on this, we use $n = 50$ as a starting point. PC maps up to this component explain $\sim 98\%$ of the total variance and we consider that it represents a balance between including too many maps, resulting in overcorrection, and discarding too many of them, so we risk not accounting for enough contaminants. Then, we obtain the \isd-PC<50 weights and we observe that these weights cause no significant overcorrection on contaminated mocks, as explained in Section \ref{sec:redmagic_discussion}. After this, we verify that the \isd -PC<50 weights show negligible levels of cross-correlation with $\kappa$, similar to those from \isd -STD34. Moreover, the recovered correlation function from these weights is in excellent agreement with that from \isd -STD34 weights, as it is shown on Figure \ref{fig:wtheta_comparison}.

In order to make the rejection of PC maps that could be causing the overcorrection as specific as possible, we cross-correlate $\kappa$ directly with the maps that contribute to the overcorrecting \isd-PC107 weights (according to the multiplicative way of \isd to make weights). However, we do not identify any individual map or family of maps clearly causing the excess correlation. In general, the PC maps that have the highest $\kappa$ correlation are the highest principal components (which have the smallest contribution to the total variance of the STD maps). Given this, we decide to test removing all PC maps above a given component. We test multiple cutoffs with PC$<n$, evaluating their clustering amplitudes, as it is shown in Fig.~\ref{fig:wtheta_pc_cutoffs}. We find that the clustering amplitudes yielded by the \isd -PC$<n$ weights with $n$ between 20 and 60 converge to similar values, while for higher $n$ it jumps abruptly to lower amplitudes. This result, together with the large amount of variance contained up to PC<50 and the impossibility of flagging a specific set of PC maps among the highest components as the culprit ones of the overcorrection, motivates the choice of $n = 50$ as our final cutoff. 

\section{Non-linear contamination with \isd }\label{app:chi2_null_dist}

\begin{figure*}
    \centering
    \includegraphics[width=\linewidth]{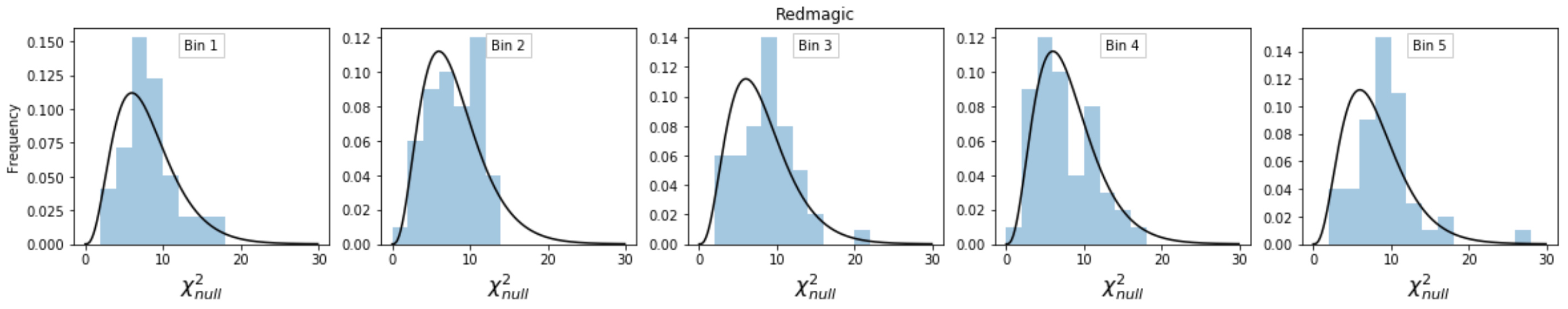}
    \caption{$\chi^2_{\rm null}$ distributions (blue histograms) for the \isd-PC<50 weighted \redmagic sample compared with a $\chi^2$ with ten degrees of freedom (black lines). Given the good agreement between both distributions, we find no clear evidence of deviations from linearity in the 1D relations of the weighted data. We find similar results for \maglim sample. }
    \label{fig:chi2_null_dist}
\end{figure*}

In order to look for non-linear contamination still present on the data after applying weights, we evaluate the distribution of $\chi^2_{\rm null}$ values from the 1D relations of the \isd -PC<50 weighted data. This kind of contamination could be undetected when using a linear model, as \isd does, and would result on high $\chi^2_{\rm null}$ values. In Figure \ref{fig:chi2_null_dist}, we show the values obtained for \redmagic. The distributions obtained for each redshift bin are not significantly different from a $\chi^2$ with ten degrees of freedom (number of 1D bins used). We obtain similar results for \maglim sample. Therefore, we find no clear evidence of the presence of non-linear contamination in our weighted data that could have been unaccounted for. 
 
\bsp	
\label{lastpage}
\end{document}